\documentclass[reprint,superscriptaddress,preprintnumbers,amsmath,amssymb,aps,prd,tightenlines,longbibliography,nofootinbib,balancelastpage]{revtex4-2}

\usepackage{graphicx}
\usepackage[dvipsnames]{xcolor}
\usepackage[sort&compress]{natbib}
\usepackage{amsmath,amssymb,bm,bbm,slashed,subdepth}
\usepackage{xr-hyper}
\usepackage[colorlinks=true
,urlcolor=blue
,anchorcolor=blue
,citecolor=blue
,filecolor=blue
,linkcolor=red
,menucolor=blue
,linktocpage=true
,pdfproducer=medialab
,pdfa=true
]{hyperref}
\usepackage{cleveref}
\usepackage{enumerate}
\usepackage{epsfig, subfigure}
\usepackage{setspace}
\usepackage{booktabs, tabularx}
\usepackage{units}
\usepackage{placeins}
\usepackage{multirow}
\usepackage{mathtools}
\usepackage[normalem]{ulem}

\newcommand{\be}{\begin{eqnarray}}
\newcommand{\ee}{\end{eqnarray}}

\newcommand{\w}{\omega}

\newcommand{\EM}{{\scriptscriptstyle \textrm{EM}}}
\newcommand{\SQ}{{\scriptscriptstyle \textrm{SQ}}}
\newcommand{\TT}{{\scriptscriptstyle \textrm{TT}}}
\newcommand{\LC}{{\scriptscriptstyle \textrm{LC}}}

\newcommand{\bB}{\mathbf{B}}
\newcommand{\bxi}{\boldsymbol{\xi}}
\newcommand{\br}{\boldsymbol{r}}
\newcommand{\bj}{\boldsymbol{j}}
\newcommand{\bdr}{\boldsymbol{\delta r}}
\newcommand{\bu}{\boldsymbol{u}}
\newcommand{\bv}{\boldsymbol{v}}
\newcommand{\bx}{\boldsymbol{x}}
\newcommand{\bfo}{\boldsymbol{f}}
\newcommand{\he}{\hat{\boldsymbol{e}}}

\usepackage[page,toc,titletoc,title]{appendix}

\begin{document}

\preprint{CERN-TH-2024-132}

\title{Magnets are Weber Bar Gravitational Wave Detectors}

\author{Valerie~Domcke}
\affiliation{Theoretical Physics Department, CERN, 1 Esplanade des Particules, CH-1211 Geneva 23, Switzerland}

\author{Sebastian~A.~R.~Ellis}
\affiliation{D\'epartement de Physique Th\'eorique, Universit\'e de Gen\`eve, 
24 quai Ernest Ansermet, 1211 Gen\`eve 4, Switzerland}

\author{Nicholas~L.~Rodd}
\affiliation{Theory Group, Lawrence Berkeley National Laboratory, Berkeley, CA 94720, USA}
\affiliation{Berkeley Center for Theoretical Physics, University of California, Berkeley, CA 94720, USA}

\begin{abstract}
When a gravitational wave (GW) passes through a DC magnetic field, it couples to the conducting wires carrying the currents which generate the magnetic field, causing them to oscillate at the GW frequency.
The oscillating currents then generate an AC component through which the GW can be detected -- thus forming a resonant mass detector or a \textit{Magnetic Weber Bar.}
We quantify this claim and demonstrate that magnets can have exceptional sensitivity to GWs over a frequency range demarcated by the mechanical and electromagnetic resonant frequencies of the system; indeed, we outline why a magnetic readout strategy can be considered an optimal Weber bar design.
The concept is applicable to a broad class of magnets, but can be particularly well exploited by the powerful magnets being deployed in search of axion dark matter, for example by DMRadio and ADMX-EFR.
Explicitly, we demonstrate that the MRI magnet that is being deployed for ADMX-EFR can achieve a broadband GW strain sensitivity of $\sim$$10^{-20}/\sqrt{\text{Hz}}$ from a few~kHz to about 10\,MHz, with a peak sensitivity down to $\sim$$10^{-22}/\sqrt{\text{Hz}}$ at a kHz exploiting a mechanical resonance.
\end{abstract}

\maketitle

The universality of the gravitational coupling implies there are many ways that a gravitational wave (GW) can interact with matter and therefore many ways GWs could be detected.
Nevertheless, the search for GWs has been historically dominated by considering the mechanical coupling of the wave; this underpins the common interpretation of Weber bars~\cite{Weber:1967jye} and interferometers~\cite{Gertsenshtein:1962kfm}, where the wave couples to a resonant mass or the interferometer mirrors, respectively.

While exploring searches for GWs at higher frequencies ($f > 1\,{\rm kHz}$), the full set of gravitational couplings is being reconsidered, as partially reviewed in Ref.~\cite{Aggarwal:2020olq}.
Two of the leading approaches are to exploit the coupling of GWs to electromagnetism~\cite{Ejlli:2019bqj,Berlin:2021txa,Domcke:2022rgu} or to again rely on the traditional mechanical coupling, but in setups optimised for short wavelength GWs, such as bulk acoustic wave devices~\cite{Goryachev:2014yra} or a levitated sensor detector (LSD)~\cite{Arvanitaki:2012cn,Aggarwal:2020umq}. 
An advantage of the mechanical coupling is that the excitations in materials are less stiff than those in electromagnetism -- the speed of sound is significantly smaller than the speed of light -- making it easier for a GW to mechanically deform or displace objects than to induce an electromagnetic (EM) field. 
A disadvantage, however, is that the induced mechanical motion then typically needs to be read out by an electromagnetic sensor, involving the need to transduce the mechanical signal to an electromagnetic one.

\begin{figure}[t]
\includegraphics[width = \columnwidth]{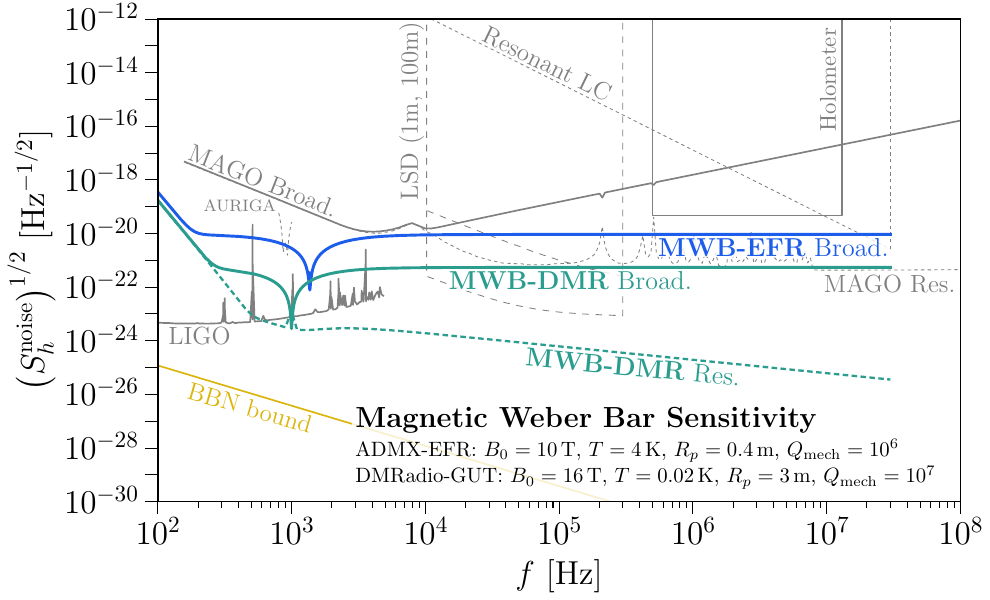}
\vspace{-0.8cm}
\caption{The noise-equivalent strain power spectral density (PSD) for two different experimental configurations with key parameters given in the legend.
The blue line corresponds roughly to the ADMX-EFR magnet~\cite{2023APS..APRC01002K} with $\ell = 2\,\text{m}$, while the teal curves correspond roughly to a magnet of the size envisioned for DMRadio-GUT~\cite{DMRadio:2022jfv} with $\ell = 4\,\text{m}$.
All solid and dashed curves correspond to broadband and resonant readouts, respectively.
See text for details.
}
\label{fig:ShPlotMagnet}
\vspace{-0.6cm}
\end{figure}

In this letter we put forward a proposal that combines the best of these two worlds. 
We consider the mechanical coupling of a GW to the support structure of a DC magnet, and in particular to the conducting wires which generate the magnetic field.
Heuristically, the GW leads to an oscillation of the wires with the GW frequency, resulting in a small AC component to the magnetic field. 
Hence, despite the coupling being mechanical, the generated signal is intrinsically electromagnetic, and can be read out through a SQUID coupled to a pickup loop, with the possibility of enhancing the signal by introducing a resonant LC circuit. 
Compared to traditional Weber Bar detectors, this greatly improves the sensitivity off the mechanical resonance frequency.
For a related discussion in the context of resonant cavities, see Ref.~\cite{Berlin:2023grv}.

The expected sensitivity of this setup is shown in Fig.~\ref{fig:ShPlotMagnet}. 
The dip around a kHz indicates a mechanical resonance of the magnet with only a single mechanical resonance shown for simplicity.
At the resonance the device behaves almost identically to a conventional Weber Bar and remarkably, in this very narrow frequency regime, the achievable strain sensitivity is comparable to that currently achieved by interferometers~\cite{PhysRevD.102.062003,Holometer:2016qoh} and can surpass the expected broadband sensitivity of the proposed MAGO experiment~\cite{Berlin:2023grv}.
The solid projected reach curves assume a broadband readout, implying that the achievable signal-to-noise ratio across the entire frequency range benefits from the full instrument integration time.
This makes the configuration highly sensitive to transients at unknown frequencies and even for persistent signals as we can exploit their duration.
This is in contrast to proposals operating in resonant mode relying on a scanning strategy (dashed gray curves)~\cite{Vinante:2006uk,Cerdonio:1997hz,Goryachev:2021zzn,Aggarwal:2020umq,Domcke:2022rgu,Berlin:2023grv}, which conventionally optimise their sensitivity over a small frequency range for a short duration before moving on.
Supplementing our proposal by coupling to a resonant circuit can improve the strain sensitivity at the corresponding resonant frequency, as we show in dashed teal.

Figure~\ref{fig:ShPlotMagnet} further displays the strain sensitivity corresponding to the cosmological bound on a stochastic gravitational wave background, labelled BBN bound~\cite{Yeh:2022heq}.
We emphasise that this comparison depends crucially on how well a given detector is suited to search for stochastic backgrounds.
Ground-based interferometers, making use of the cross-correlation across different detectors,  currently reach sensitivities about two orders of magnitude beyond this limit.
More generally, broadband detectors can rely on templates (often simple power laws) for the expected signal spectra to improve the sensitivity by a factor $(t_\text{int} \Delta f)^{1/4}$, with $t_\text{int}$ the integration time and $\Delta f$ the minimum of the signal and analysis bandwidths.

The goal of this letter is to provide a careful analysis of the claims made so far.
We begin, however, with a sketch of the essential idea.

\vspace{0.2cm}
\noindent {\bf Wiggling a DC Magnet.}
%
Consider a DC solenoidal magnet of length $\ell$ made of $N$ coils carrying current $I$.
Unperturbed, it generates a field $B_0 \sim N I/\ell$.
Working in the local inertial frame associated with an observer in the laboratory, a passing GW of strain $h$ imparts the equivalent of a Newtonian force on the experiment, slightly deforming its shape.
The exact deformation depends on the material properties and geometry of the solenoid.
However, if we consider a long thin solenoid and frequencies well above the mechanical resonance,  a GW orthogonal to the symmetry axis of the solenoid induces a deformation that is approximately $\ell \to \ell + h \ell$.
The solenoid now generates a magnetic field of $B \sim B_0 (1-h)$, implying the presence of an AC contribution to the magnetic field oscillating at the GW frequency.

To read out the signal we place a pickup loop of radius $R_p$ through which the GW induces an AC flux of $\Phi \sim h B_0 \pi R_p^2$.
Imagining a broadband readout, we can estimate our sensitivity by inductively coupling the pickup loop to a SQUID.
Taking a characteristic coupling of $\kappa \sim 10^{-2}$ and noise of $\Phi_\SQ \sim 10^{-21}\,{\rm Wb}/\sqrt{\rm Hz}$, the sensitivity expressed as a noise spectral densities is $(S_h^{\rm noise})^{1/2} \sim  \Phi_\SQ/(\kappa B_0 \pi R_p^2) \sim 10^{-20}/\sqrt{\rm Hz}$ for $R_p \sim 0.4\,{\rm m}$ and $B_0 \sim 10\,{\rm T}$, consistent with Fig.~\ref{fig:ShPlotMagnet}.
For a persistent monochromatic GW signal, the strain sensitivity of a ten day observation would be $h \sim (S_h^{\rm noise}/T)^{1/2} \sim 10^{-23}$, which could probe GWs emitted by axion superradiance at around 0.1\,MHz, an idea which has been broadly explored, see e.g. Refs.~\cite{Ternov:1978gq,Zouros:1979iw,Detweiler:1980uk,Arvanitaki:2009fg,Arvanitaki:2010sy,Arvanitaki:2012cn,Yoshino:2013ofa,Brito:2014wla,Arvanitaki:2014wva,Brito:2015oca,Brito:2017zvb,Tsukada:2018mbp,Aggarwal:2020umq,Zhu:2020tht,Brzeminski:2024drp}.
We validate the accuracy of these heuristic estimates of the signal and noise in the following sections.

Before doing so, let us briefly justify where our concept improves over existing approaches.
Comparing with traditional Weber bars, the main advantage lies in inherently large EM energy in a magnetic readout, enabling an efficient measurement of the mechanical deformation.
Comparing to a direct coupling of the GW to the EM field, we profit from the reduced stiffness of the mechanical deformations at frequencies below the EM resonance.
Both comparisons are discussed in greater detail in the Supplemental Material (SM), which
includes Refs.~\cite{Kahn:2016aff,Bonaldi:2003ah,PhysRevLett.100.227006,kumar2016origin,rower2023evolution,Kubo:1966fyg,Saulson:1990jc,Foster:2017hbq,Benabou:2022qpv}.

\vspace{0.2cm}
\noindent {\bf Induced magnetic field.}
%
Throughout this letter we work in the proper detector frame, in which the GW acts by exerting a Newtonian force
\begin{align}
F^h_i/m = \frac{1}{2} \ddot{h}_{ij}^\TT \br^j,
\label{eq:force}
\end{align}
acting on a test particle of mass $m$ at position $\br$, with $h_{ij}^\TT$ indicating the GW tensor evaluated in the transverse traceless frame, which depends on the amplitude for the two polarisations $h_+$ and $h_{\times}$.
Explicitly, we decompose a GW propagating along the $x$-axis as $h_{ij}^\TT(t) = h_A e_{ij}^A e^{-i\omega (t-x)}$, where $A=+,\times$ and $e_{ij}^A$ are the polarisation tensors, with explicit forms given in the SM.
We consider the impact of this force on a solenoidal magnetic field, generated by a current-carrying spool.
In detail, the force will perturb both the position and orientation of the spool through which the currents are running.
As we show in the SM, applying the Biot-Savart law yields
\be
\bB(\br') = \int_V d^3\br\, \frac{\bj(\br) \times (\br' - \bxi(\br))}{4\pi |\br' - \bxi(\br)|^3}.
\label{eq:B_master}
\ee
The volume integral is performed over the unperturbed (flat space) coordinates of the spool.
The deformation of the spool's location by the GW is encoded in $\bxi(\br) = \br + \bdr(\br)$ and we discuss how this is determined shortly.
Henceforth we leave the dependence on $\br$ implicit.
The current density is given by
\be
\bj = \frac{I_N}{\ell \, \Delta r} \frac{d \bxi}{d \phi} \left| \frac{d \bxi}{d \phi} \right|^{-1}\!,
\ee
with $I_N = N I$ denoting the ampere-turns of the current, $\Delta r$ the radial width of the spool, $\ell$ its length, and the remaining factors indicating the orientation of the current in the presence of a GW.

Equation~\eqref{eq:B_master} demonstrates that the GW modifies the magnetic field generated by the spool, adding an oscillatory component at the GW frequency.
We can read this out with a pickup loop that is sensitive to the field at all positions $\br'$ within its area.
In computing this effect, one must account for the fact the pickup loop itself is subject to the GW force.
In the following, we will assume the pickup loop to be suspended, so that we can treat its motion as approximately free-falling, i.e.\ the GW perturbs the flat space position $\br'_0$ to $\br' = \br'_0 + \bdr'_\text{ff}$.
We do not consider possible mechanical resonances of the pickup loop, as with the geometry envisioned, the dominant effect is well captured in this approximation.
In that limit, the deformation of the pickup loop is simply given by the Newtonian force of Eq.~\eqref{eq:force},
\be
\ddot{\bdr}'_{\text{ff},i} = \frac{1}{2} \ddot h_{ij}^\TT \br_0^{\prime j}.
\label{eq:FF}
\ee
We note that the above treatment cannot be extended to arbitrarily low frequencies.
Once the GW frequency falls below the restoring frequency of the suspensions system employed for the pickup loop, the magnet and loop would begin to move in concert, suppressing our sensitivity.

What remains is to determine the deformation of the spool $\bdr$.
We consider two approaches.
Well above the resonant frequencies of the spool, the GW acts as a driving frequency the material cannot respond to in time, implying we can take the free-falling limit with $\bdr$ determined from Eq.~\eqref{eq:FF}.
This corresponds to the molecules of the spool material responding individually to the GW.
At lower frequencies, we must account for the response of the material to the imposed force.
We do so using the mechanical eigenmodes of the magnet parameterised by dimensionless displacements $\bu_{mnp}(\br)$, with $(m,n,p)$ labelling the order of the modes in the polar, radial, and longitudinal directions.
Details of how these modes can be approximated for a finite width cylinder are provided in the SM.
The mechanical response of the magnet to an incoming GW is parameterised in terms of dimensionless overlap factors,
\be
\eta_{mnp}^A = \frac{1}{2V \ell_{\eta}} \int_V d^3 \br\, e_{ij}^A \br^i[\bu_{mnp}^*]^j,
\label{eq:overlap}
\ee
normalised such that if we sent $[\bu_{mnp}^*]^j \to \tfrac{1}{2} e_{ij}^A \br^j/\ell_\eta$ for either polarisation we obtain $\eta = 1$.
As an example to demonstrate that ${\cal O}(1)$ couplings are possible, consider a mode that we expect to strongly couple to a GW traveling along the $z$-axis, $\bu_{210}$ (the explicit form is given in the SM).
If we take a spool of length 2\,m and inner and outer radii of 0.6\,m and 0.65\,m, we find a coupling of $\eta^{+,\times}_{210} \simeq 0.95$, with the + ($\times$) polarisation of the GW coupling to the 210 mode with odd (even) azimuthal dependence.
A GW incident along the $x$-axis will also couple to this mode, albeit with a reduced coupling of $\eta_{210}^+ \simeq 0.41$ and $\eta_{210}^\times \simeq 0$.
The magnet dimensions are inspired by the MRI magnet to be used for ADMX-EFR, which has a peak magnetic field of $B_0 \simeq 10\,\text{T}$ generated by a current of $\sim 20$ MA-turns.
The eigenfrequency of this mode is determined by the diameter of the spool, and for the dimensions above is found to be at $f \simeq 1.4$\,kHz for a stainless steel cylinder. 
To compute the induced magnetic field we insert $\bdr = \eta_{210} \bu_{210} \ell_\eta$ into $\bxi$, which can be enhanced by the mechanical quality factor $Q_\text{mech}$ on resonance, in Eq.~\eqref{eq:B_master} and otherwise proceed as in the freely-falling limit.

The precise magnitude of the signal depends on the position and orientation of the pickup loop, and we provide numerical results for different frequency regimes in the SM.
For example, one can obtain a large signal by placing a pickup loop close to an end cap of the magnet, where the external magnetic field is still rather strong but features a significant gradient.
With a suitable placement of the pickup loop, we expect induced magnetic fields of ${\cal O}(h B_0)$ at frequencies away from the mechanical resonance and ${\cal O}(h Q_\text{mech} B_0)$ on the mechanical resonance.
In a realistic setup, material simulations and measurements will be needed to determine the precise mechanical response of the system.

\vspace{0.2cm}
\noindent {\bf Signal-to-Noise Ratio (SNR).}
%
The optimal SNR for the strain $h$ can be written as\footnote{As written, this SNR is quadratic in $h$.
For a signal whose waveform can be matched in the time domain, an SNR that is linear in strain can be constructed by performing a matched filtering analysis.
Doing so leads to an SNR that is identical to Eq.~\eqref{eq:SNR}, only with the frequency integral taken over the ratio of un-squared PSDs.}
\be
\text{SNR}^2 \simeq 2 t_{\rm int} \int_0^{\infty} df \left(\frac{S_{\rm sig}(f)}{S_{\rm noise}(f)}\right)^2\!.
\label{eq:SNR}
\ee
As derived explicitly in the SM, the signal power spectral density (PSD) enters in the form of a PSD of the flux through the pickup loop.
Explicitly, $S_{\rm sig} (f) = B_0^2 A_p^2 |\mathcal{G}(f,\bx)|^2 S_h(f)$, where $A_p$ is the pickup loop area, $S_h(f)$ is the GW PSD, whilst $\mathcal{G}(f,\bx)$ is a dimensionless gain factor that depends on the GW frequency as well as the position and orientation of the pickup loop and which must be computed numerically.
Lastly, $S_{\rm noise}(f)$ represents the PSD of various noise sources that we enumerate in the SM.

The dimensionless gain factor arises from the two contributions to the magnetic flux that traverses the pickup loop, as discussed in the previous section.
The first is the coupling of the GW to the spool.
Numerically, we conservatively only include the gain generated from the GW coupling to the $210$ mode as discussed above; in principle there are contributions from all modes to which the GW couples, although this mode will dominate for a GW propagating along the $z$ direction.
The second contribution comes from the relative motion of the freely-falling pickup loop with respect to the magnet.

Let us consider the expected behaviour of the gain in three limits around the lowest-lying mechanical resonance of the spool $f_{\rm min}$: 1. $f \ll f_{\rm min}$; 2. $f \sim f_{\rm min}$; and 3. $f \gg f_{\rm min}$.
In the first case, the low frequency limit, the mechanical coupling to the spool is suppressed by $(f/f_\text{min})^2$, whereas the coupling to the freely falling pickup loop is not.
Consequently, the latter dominates, leaving $|\mathcal{G}|^2_{f\ll f_{\rm min}} \lesssim 1$.
In this regime, the SQUID noise is likely to dominate in a broadband setup.
As a result, if both the signal and noise PSDs can be approximated as flat in a bandwidth $\Delta f$, the SNR for this regime can be obtained from Eq.~\eqref{eq:SNR} as
\be
\text{SNR}_{\rm f\ll f_{\rm min}} \sim \left(t_{\rm int} \Delta f\right)^{1/2} \frac{S_h(f)\, B_0^2\,A_{p}^2}{S_{\rm noise}^{\SQ}(f)}.
\label{eq:SNR_lowF_broadband}
\ee

In the second regime, where the GW frequency matches a mechanical resonance $f\sim f_{\rm min}$, the gain is dominated by the mechanical quality factor moderating the response of that mode.
The gain again depends on the positioning of the pickup loop (see SM), although largest it can be is $|\mathcal{G}|^2 \sim Q_{\rm mech}^2$.
In this case, thermal mechanical noise in the magnet dominates for typical parameter choices, in which case one would not gain further with a resonant rather than broadband EM readout (as seen in Fig.~\ref{fig:ShPlotMagnet}).
(We note that thermal noise in the pickup loop remains subdominant at all frequencies.)
The resulting SNR scales as
\be
\hspace{-0.6cm}
\text{SNR}_{\rm mech.\,res.} \sim \left(t_{\rm int} \Delta f\right)^{1/2} \frac{Q_{\rm mech}\,S_h(f)\,\ell_h^2\,M\,\w_{\rm mech}^3}{2\,T},
\label{eq:SNR_mech_res}
\ee
where $M$ is the mass of the magnet, $T$ its temperature, and the characteristic scale of the deformations induced by the GW is $\ell_h = \eta \ell_\eta$, written in terms of the overlap factor in Eq.~\eqref{eq:overlap}.\footnote{The reader should be cautioned that taking the $M\to\infty$ limit does not infinitely improve the SNR.
For a broadband readout, eventually SQUID noise again becomes the dominant background and the SNR saturates at $Q_{\rm mech}^2\times \text{SNR}_{\rm f\ll f_{\rm min}}$.}
In this thermally-limited mechanically-resonant scenario, the broadband SNR scales as $Q_{\rm mech}$ owing to the suppression of the thermal mechanical noise by the mechanical resonance's linewidth, $\w_{\rm mech}/Q_{\rm mech}$.

Last, we turn to the limit where $f \gg f_{\rm min}$.
Now we are in the flexible regime discussed previously, and the gain is approximately $|\mathcal{G}|^2  \lesssim 1$.
In this regime, the $\bu_{210}$ mode responds flexibly, and the sum over all modes would approximate the free-falling magnet limit.
SQUID noise is expected to dominate, and the SNR is again approximated by Eq.~\eqref{eq:SNR_lowF_broadband}.

In all three regimes discussed above, we have assumed that the pickup loop is placed just outside the magnet.
Flux conservation through the superconducting magnet coils implies that for $f \gg f_{\rm min}$ there will be no signal through a coaxial loop placed inside the magnet.

Note that the noise sources we consider here will dominate as long as there is sufficient shielding against ambient magnetic field noise.
Given the anticipated magnetic field noise of $B(f) \lesssim 10^{-14}\,\text{T}/\sqrt{\text{Hz}}$ at frequencies $f\gtrsim 100\,\text{Hz}$~\cite{CONSTABLE2023107090}, the shielding requires a reduction of at most $10^7$ in the field amplitude.
This could be achieved, for example, with a 5mm-thick iron-nickel alloy shield with $\mu_r = 75\times10^3,~\sigma = 2\times10^6\,\text{S/m}$~\cite{celozzi2023electromagnetic}.

\vspace{0.2cm}
\noindent {\bf Sensitivity.}
%
The expressions from the previous section can be used to determine the sensitivity of a single detector to both stochastic GW sources, where the bin width $\Delta f$ depends on the expected spectral shape of the GW signal, and to coherent sources for which one should take $S_h(f) \sim h_0^2/\Delta f$.
The information can also be used to determine the expected sensitivity of the detector as encoded in the noise-equivalent strain, which is shown in Fig.~\ref{fig:ShPlotMagnet}.
Formally, the noise-equivalent strain PSD is defined as $S_h^{\rm noise}(f) \equiv S_{\rm noise}(f) S_h(f)/S_{\rm sig}(f)$.
Detailed expressions for each component are provided in the SM.

We can already estimate the form of $S_h^{\rm noise}(f)$ from the details of the previous section.
In particular, taking the form for the signal PSD discussed below Eq.~\eqref{eq:SNR}, we have $S_h^{\rm noise}(f) \sim S_{\rm noise}(f)/B_0^2 A_p^2 |\mathcal{G}(f,\bx)|^2$, exposing its various dependencies.
In Fig.~\ref{fig:ShPlotMagnet} we assume a gain of $|\mathcal{G}|^2 = 5$ away from a mechanical resonance, whereas on the resonance we directly compute the sensitivity by comparing the displacement PSDs as detailed in the SM.
Combining these two approaches we find a slight degradation in the sensitivity at the mechanical resonance.
These specific sensitivities correspond to the values found for the single mode studied in the SM, so it is likely that a detailed numerical study including all modes and optimising the loop placement can improve upon these.
In particular, we anticipate the possibility of improvement around a mechanical resonance of a mode with a longitudinal component.
Of course, the additional modes would also enhance the thermal noise at their respective locations, which we have neglected in our projections.

Finally, coupling the pickup loop to an LC circuit with EM quality factor $Q_{\EM}$ can improve the sensitivity in parts of the parameter space at the cost of reducing the detector bandwidth.
As a simple estimate, on the mechanical resonance the SNR would remain the same as in the broadband case of Eq.~\eqref{eq:SNR_mech_res} (and therefore again the sensitivity slightly degrades), whereas above the mechanical resonance, the SNR would improve from Eq.~\eqref{eq:SNR_lowF_broadband} by a factor $\sim 10^5 \times(f/10^4\,\text{Hz})$ assuming resonant LC circuit parameters of $T_{\LC} = 10\,\text{mK}$ and $Q_{\EM}=2 \times 10^7$, comparable to DMRadio-GUT~\cite{DMRadio:2022jfv}.
The intrinsic bandwidth of the LC resonator $\Delta f_{\rm LC}$ can affect the determination of $\Delta f$ in the resulting expressions for the SNR. 
For more details, see the SM.

In Fig.~\ref{fig:ShPlotMagnet}, we have taken two hypothetical magnet and pickup loop configurations, inspired by the ADMX-EFR and DMRadio-GUT magnets. 
The main parameters we have chosen for both are described in the figure caption.
For the ADMX-EFR-inspired scenario, we computed the resonant frequency of the $u_{210}$ mode, and assumed it has a quality factor of $Q_{\rm mech} = 10^6$.
For the DMRadio-GUT-inspired configuration, we assume $Q_{\rm mech} = 10^7$ and a resonance at $f_{\rm min} =1\,\text{kHz}$, justified by the large magnet dimensions.
For ADMX-EFR we assumed a similar magnet temperature as achieved for AURIGA of 4\,K, whereas for DMRadio-GUT we adopted a more aggressive 0.02\,K, although this only impacts the sensitivity on the mechanical resonance.
We conservatively assume that both magnets weigh $M = 40\,\text{tons}$ (similar to previous resonant mass experiments~\cite{PhysRevD.75.022002}), such that any increase in the DMRadio-GUT magnet mass could be absorbed by an overestimate in the mechanical quality factor.
For simplicity we assume persistent superconducting magnets, otherwise additional noise associated with the external power supply must be included.
Two pickup loop configurations were employed: A co-axial loop (off-resonance sensitivity) and a quarter-circle loop (sensitivity on resonance), both placed close to the endcap of the magnet.
The sensitivity further assumes seismic isolation of the apparatus through a dual suspension system assuming a quiet site~\cite{Saulson:2017jlf}, which is eventually overcome leading to a loss in sensitivity at low frequencies.
Even in the case of full seismic isolation, the suspension system is likely to contribute to a deterioration of the sensitivity below $f\lesssim 100\,\text{Hz}$, a frequency associated to the typical size of an experimental hall, and eventually gravity gradient noise~\cite{Saulson:1984yg,Hughes:1998pe} will become relevant.

\vspace{0.2cm}
\noindent {\bf Conclusions.}
%
The focus of this letter has been to demonstrate that DC magnets can act as remarkably sensitive GW detectors.
The mechanical force exerted by a GW on the magnet itself and on a pickup loop placed within the magnetic field directly induces an AC magnetic flux component through the pickup loop.
Our estimate for the resulting sensitivity is shown in Fig.~\ref{fig:ShPlotMagnet}; our projected noise-equivalent strain with a broadband readout is stronger than many projections achieve with a narrow band resonant readout.
Our more conservative estimate is based on the magnet dimensions suggested for ADMX-EFR~\cite{2023APS..APRC01002K}, and can be easily generalised to other powerful magnets, such as the GrAHal magnet~\cite{Grenet:2021vbb} or the magnets envisioned for DMRadio~\cite{DMRadio:2022jfv,DMRadio:2022pkf,DMRadio:2023igr}.
Away from the mechanical resonance, our sensitivity could also be improved with the use of a resonant EM readout.

Looking forward, the observations of this letter will impact GW searches relying on static magnetic fields.
The most obvious connection is to searches inspired by low-mass axion haloscopes~\cite{Domcke:2022rgu}.
For toroidal magnets (as used for ABRA-10cm~\cite{Ouellet:2018beu,Ouellet:2019tlz,Salemi:2021gck} or SHAFT~\cite{Gramolin:2020ict}) the effect discussed here is irrelevant as the pickup loop is typically placed in a field free region.
On the contrary, for solenoidal magnets (as used in BASE~\cite{Devlin:2021fpq}, WISPLC~\cite{Zhang:2021bpa}, ADMX SLIC~\cite{Crisosto:2019fcj}, and proposed for DMRadio-m$^3$~\cite{DMRadio:2022pkf,DMRadio:2023igr}), both the contribution discussed here and the contribution from the EM coupling~\cite{Domcke:2023bat} are present.
The details depend on the placement of the pickup loop, but as discussed, we expect the mechanical signal to dominate below the EM resonant frequency.
As a general statement, our work is an explicit realisation of the fact that in the presence of a GW, any laboratory magnetic field cannot be considered static.
This is particularly relevant at frequencies around the mechanical resonances, where the magnet can be treated neither as rigid (low frequency limit) or as free-falling (high frequency limit), see also Ref.~\cite{Ratzinger:2024spd}.

\vspace{0.2cm}
\noindent {\it Acknowledgements.}
%
We thank Krisztian Peters and J\'er\'emie Quevillon for helpful feedback. 
We particularly thank Aaron Chou for helpful feedback regarding flux conservation through a superconducting wire, and Andrew Sonnenschein for discussions and providing details regarding the ADMX-EFR magnet.
We further thank Kaliro\"e Pappas and Jan Sch\"utte-Engel for useful comments on a draft version of this work.
The work of SARE was supported by SNF Ambizione grant PZ00P2\_193322, \textit{New frontiers from sub-eV to super-TeV}.
The work of NLR was supported by the Office of High Energy Physics of the U.S. Department of Energy under contract DE-AC02-05CH11231.
Part of this work was performed at Aspen Center for Physics, which is supported by National Science Foundation grant PHY-2210452.

\bibliographystyle{utphys}
\bibliography{Biblio}

\providecommand{\href}[2]{#2}\begingroup\raggedright\begin{thebibliography}{10}

\bibitem{Weber:1967jye}
J.~Weber, ``{Gravitational Radiation},''
  \href{http://dx.doi.org/10.1103/PhysRevLett.18.498}{{\em Phys. Rev. Lett.}
  {\bf 18} (1967) no.~13, 498}.

\bibitem{Gertsenshtein:1962kfm}
M.~E. Gertsenshtein and V.~I. Pustovoit, ``{On the Detection of Low Frequency
  Gravitational Waves},'' {\em Sov. Phys. JETP} {\bf 16} (1962)  433.

\bibitem{Aggarwal:2020olq}
N.~Aggarwal {\em et al.}, ``{Challenges and opportunities of gravitational-wave
  searches at MHz to GHz frequencies},''
  \href{http://dx.doi.org/10.1007/s41114-021-00032-5}{{\em Living Rev. Rel.}
  {\bf 24} (2021) no.~1, 4}, \href{http://arxiv.org/abs/2011.12414}{{\tt
  arXiv:2011.12414 [gr-qc]}}.

\bibitem{Ejlli:2019bqj}
A.~Ejlli, D.~Ejlli, A.~M. Cruise, G.~Pisano, and H.~Grote, ``{Upper limits on
  the amplitude of ultra-high-frequency gravitational waves from graviton to
  photon conversion},''
  \href{http://dx.doi.org/10.1140/epjc/s10052-019-7542-5}{{\em Eur. Phys. J. C}
  {\bf 79} (2019) no.~12, 1032}, \href{http://arxiv.org/abs/1908.00232}{{\tt
  arXiv:1908.00232 [gr-qc]}}.

\bibitem{Berlin:2021txa}
A.~Berlin, D.~Blas, R.~Tito~D'Agnolo, S.~A.~R. Ellis, R.~Harnik, Y.~Kahn, and
  J.~Sch\"utte-Engel, ``{Detecting high-frequency gravitational waves with
  microwave cavities},''
  \href{http://dx.doi.org/10.1103/PhysRevD.105.116011}{{\em Phys. Rev. D} {\bf
  105} (2022) no.~11, 116011}, \href{http://arxiv.org/abs/2112.11465}{{\tt
  arXiv:2112.11465 [hep-ph]}}.

\bibitem{Domcke:2022rgu}
V.~Domcke, C.~Garcia-Cely, and N.~L. Rodd, ``{Novel Search for High-Frequency
  Gravitational Waves with Low-Mass Axion Haloscopes},''
  \href{http://dx.doi.org/10.1103/PhysRevLett.129.041101}{{\em Phys. Rev.
  Lett.} {\bf 129} (2022) no.~4, 041101},
  \href{http://arxiv.org/abs/2202.00695}{{\tt arXiv:2202.00695 [hep-ph]}}.

\bibitem{Goryachev:2014yra}
M.~Goryachev and M.~E. Tobar, ``{Gravitational Wave Detection with High
  Frequency Phonon Trapping Acoustic Cavities},''
  \href{http://dx.doi.org/10.1103/PhysRevD.90.102005}{{\em Phys. Rev. D} {\bf
  90} (2014) no.~10, 102005}, \href{http://arxiv.org/abs/1410.2334}{{\tt
  arXiv:1410.2334 [gr-qc]}}. [Erratum: Phys.Rev.D 108, 129901 (2023)].

\bibitem{Arvanitaki:2012cn}
A.~Arvanitaki and A.~A. Geraci, ``{Detecting high-frequency gravitational waves
  with optically-levitated sensors},''
  \href{http://dx.doi.org/10.1103/PhysRevLett.110.071105}{{\em Phys. Rev.
  Lett.} {\bf 110} (2013) no.~7, 071105},
  \href{http://arxiv.org/abs/1207.5320}{{\tt arXiv:1207.5320 [gr-qc]}}.

\bibitem{Aggarwal:2020umq}
N.~Aggarwal, G.~P. Winstone, M.~Teo, M.~Baryakhtar, S.~L. Larson, V.~Kalogera,
  and A.~A. Geraci, ``{Searching for New Physics with a Levitated-Sensor-Based
  Gravitational-Wave Detector},''
  \href{http://dx.doi.org/10.1103/PhysRevLett.128.111101}{{\em Phys. Rev.
  Lett.} {\bf 128} (2022) no.~11, 111101},
  \href{http://arxiv.org/abs/2010.13157}{{\tt arXiv:2010.13157 [gr-qc]}}.

\bibitem{2023APS..APRC01002K}
S.~{Knirck} and {ADMX Collaboration Team}, ``{ADMX Extended Frequency Range
  (EFR): Searching for 2-4GHz axions with 18 cavities},'' in {\em APS April
  Meeting Abstracts}, vol.~2023 of {\em APS Meeting Abstracts}, p.~CCC01.002.
\newblock 2023.

\bibitem{DMRadio:2022jfv}
{\bf DMRadio} Collaboration, L.~Brouwer {\em et al.}, ``{Proposal for a
  definitive search for GUT-scale QCD axions},''
  \href{http://dx.doi.org/10.1103/PhysRevD.106.112003}{{\em Phys. Rev. D} {\bf
  106} (2022) no.~11, 112003}, \href{http://arxiv.org/abs/2203.11246}{{\tt
  arXiv:2203.11246 [hep-ex]}}.

\bibitem{Berlin:2023grv}
A.~Berlin, D.~Blas, R.~Tito~D'Agnolo, S.~A.~R. Ellis, R.~Harnik, Y.~Kahn,
  J.~Sch\"utte-Engel, and M.~Wentzel, ``{Electromagnetic cavities as mechanical
  bars for gravitational waves},''
  \href{http://dx.doi.org/10.1103/PhysRevD.108.084058}{{\em Phys. Rev. D} {\bf
  108} (2023) no.~8, 084058}, \href{http://arxiv.org/abs/2303.01518}{{\tt
  arXiv:2303.01518 [hep-ph]}}.

\bibitem{PhysRevD.102.062003}
A.~Buikema {\em et al.}, ``Sensitivity and performance of the advanced ligo
  detectors in the third observing run,''
  \href{http://dx.doi.org/10.1103/PhysRevD.102.062003}{{\em Phys. Rev. D} {\bf
  102} (2020) no.~27, 062003}.

\bibitem{Holometer:2016qoh}
{\bf Holometer} Collaboration, A.~S. Chou {\em et al.}, ``{MHz Gravitational
  Wave Constraints with Decameter Michelson Interferometers},''
  \href{http://dx.doi.org/10.1103/PhysRevD.95.063002}{{\em Phys. Rev. D} {\bf
  95} (2017) no.~6, 063002}, \href{http://arxiv.org/abs/1611.05560}{{\tt
  arXiv:1611.05560 [astro-ph.IM]}}.

\bibitem{Vinante:2006uk}
{\bf AURIGA} Collaboration, A.~Vinante, ``{Present performance and future
  upgrades of the AURIGA capacitive readout},''
  \href{http://dx.doi.org/10.1088/0264-9381/23/8/S14}{{\em Class. Quant. Grav.}
  {\bf 23} (2006)  S103}.

\bibitem{Cerdonio:1997hz}
M.~Cerdonio {\em et al.}, ``{The Ultracryogenic gravitational wave detector
  AURIGA},'' \href{http://dx.doi.org/10.1088/0264-9381/14/6/016}{{\em Class.
  Quant. Grav.} {\bf 14} (1997)  1491}.

\bibitem{Goryachev:2021zzn}
M.~Goryachev, W.~M. Campbell, I.~S. Heng, S.~Galliou, E.~N. Ivanov, and M.~E.
  Tobar, ``{Rare Events Detected with a Bulk Acoustic Wave High Frequency
  Gravitational Wave Antenna},''
  \href{http://dx.doi.org/10.1103/PhysRevLett.127.071102}{{\em Phys. Rev.
  Lett.} {\bf 127} (2021) no.~7, 071102},
  \href{http://arxiv.org/abs/2102.05859}{{\tt arXiv:2102.05859 [gr-qc]}}.

\bibitem{Yeh:2022heq}
T.-H. Yeh, J.~Shelton, K.~A. Olive, and B.~D. Fields, ``{Probing physics beyond
  the standard model: limits from BBN and the CMB independently and
  combined},'' \href{http://dx.doi.org/10.1088/1475-7516/2022/10/046}{{\em
  JCAP} {\bf 10} (2022)  046}, \href{http://arxiv.org/abs/2207.13133}{{\tt
  arXiv:2207.13133 [astro-ph.CO]}}.

\bibitem{Ternov:1978gq}
I.~M. Ternov, V.~R. Khalilov, G.~A. Chizhov, and A.~B. Gaina, ``{Finite motion
  of massive particles in the Kerr and Schwarzschild fields},''
  \href{http://dx.doi.org/10.1007/BF00894575}{{\em Sov. Phys. J.} {\bf 21}
  (1978)  1200}.

\bibitem{Zouros:1979iw}
T.~J.~M. Zouros and D.~M. Eardley, ``{INSTABILITIES OF MASSIVE SCALAR
  PERTURBATIONS OF A ROTATING BLACK HOLE},''
  \href{http://dx.doi.org/10.1016/0003-4916(79)90237-9}{{\em Annals Phys.} {\bf
  118} (1979)  139}.

\bibitem{Detweiler:1980uk}
S.~L. Detweiler, ``{KLEIN-GORDON EQUATION AND ROTATING BLACK HOLES},''
  \href{http://dx.doi.org/10.1103/PhysRevD.22.2323}{{\em Phys. Rev. D} {\bf 22}
  (1980)  2323}.

\bibitem{Arvanitaki:2009fg}
A.~Arvanitaki, S.~Dimopoulos, S.~Dubovsky, N.~Kaloper, and J.~March-Russell,
  ``{String Axiverse},''
  \href{http://dx.doi.org/10.1103/PhysRevD.81.123530}{{\em Phys. Rev. D} {\bf
  81} (2010)  123530}, \href{http://arxiv.org/abs/0905.4720}{{\tt
  arXiv:0905.4720 [hep-th]}}.

\bibitem{Arvanitaki:2010sy}
A.~Arvanitaki and S.~Dubovsky, ``{Exploring the String Axiverse with Precision
  Black Hole Physics},''
  \href{http://dx.doi.org/10.1103/PhysRevD.83.044026}{{\em Phys. Rev. D} {\bf
  83} (2011)  044026}, \href{http://arxiv.org/abs/1004.3558}{{\tt
  arXiv:1004.3558 [hep-th]}}.

\bibitem{Yoshino:2013ofa}
H.~Yoshino and H.~Kodama, ``{Gravitational radiation from an axion cloud around
  a black hole: Superradiant phase},''
  \href{http://dx.doi.org/10.1093/ptep/ptu029}{{\em PTEP} {\bf 2014} (2014)
  043E02}, \href{http://arxiv.org/abs/1312.2326}{{\tt arXiv:1312.2326
  [gr-qc]}}.

\bibitem{Brito:2014wla}
R.~Brito, V.~Cardoso, and P.~Pani, ``{Black holes as particle detectors:
  evolution of superradiant instabilities},''
  \href{http://dx.doi.org/10.1088/0264-9381/32/13/134001}{{\em Class. Quant.
  Grav.} {\bf 32} (2015) no.~13, 134001},
  \href{http://arxiv.org/abs/1411.0686}{{\tt arXiv:1411.0686 [gr-qc]}}.

\bibitem{Arvanitaki:2014wva}
A.~Arvanitaki, M.~Baryakhtar, and X.~Huang, ``{Discovering the QCD Axion with
  Black Holes and Gravitational Waves},''
  \href{http://dx.doi.org/10.1103/PhysRevD.91.084011}{{\em Phys. Rev. D} {\bf
  91} (2015) no.~8, 084011}, \href{http://arxiv.org/abs/1411.2263}{{\tt
  arXiv:1411.2263 [hep-ph]}}.

\bibitem{Brito:2015oca}
R.~Brito, V.~Cardoso, and P.~Pani, ``{Superradiance}: {New Frontiers in Black
  Hole Physics},'' \href{http://dx.doi.org/10.1007/978-3-319-19000-6}{{\em
  Lect. Notes Phys.} {\bf 906} (2015)  1},
  \href{http://arxiv.org/abs/1501.06570}{{\tt arXiv:1501.06570 [gr-qc]}}.

\bibitem{Brito:2017zvb}
R.~Brito, S.~Ghosh, E.~Barausse, E.~Berti, V.~Cardoso, I.~Dvorkin, A.~Klein,
  and P.~Pani, ``{Gravitational wave searches for ultralight bosons with LIGO
  and LISA},'' \href{http://dx.doi.org/10.1103/PhysRevD.96.064050}{{\em Phys.
  Rev. D} {\bf 96} (2017) no.~6, 064050},
  \href{http://arxiv.org/abs/1706.06311}{{\tt arXiv:1706.06311 [gr-qc]}}.

\bibitem{Tsukada:2018mbp}
L.~Tsukada, T.~Callister, A.~Matas, and P.~Meyers, ``{First search for a
  stochastic gravitational-wave background from ultralight bosons},''
  \href{http://dx.doi.org/10.1103/PhysRevD.99.103015}{{\em Phys. Rev. D} {\bf
  99} (2019) no.~10, 103015}, \href{http://arxiv.org/abs/1812.09622}{{\tt
  arXiv:1812.09622 [astro-ph.HE]}}.

\bibitem{Zhu:2020tht}
S.~J. Zhu, M.~Baryakhtar, M.~A. Papa, D.~Tsuna, N.~Kawanaka, and H.-B.
  Eggenstein, ``{Characterizing the continuous gravitational-wave signal from
  boson clouds around Galactic isolated black holes},''
  \href{http://dx.doi.org/10.1103/PhysRevD.102.063020}{{\em Phys. Rev. D} {\bf
  102} (2020) no.~6, 063020}, \href{http://arxiv.org/abs/2003.03359}{{\tt
  arXiv:2003.03359 [gr-qc]}}.

\bibitem{Brzeminski:2024drp}
D.~Brzeminski, A.~Hook, J.~Huang, and C.~Ristow, ``{Searching for String
  Bosenovas with Gravitational Wave Detectors},''
  \href{http://arxiv.org/abs/2407.18991}{{\tt arXiv:2407.18991 [hep-ph]}}.

\bibitem{Kahn:2016aff}
Y.~Kahn, B.~R. Safdi, and J.~Thaler, ``{Broadband and Resonant Approaches to
  Axion Dark Matter Detection},''
  \href{http://dx.doi.org/10.1103/PhysRevLett.117.141801}{{\em Phys. Rev.
  Lett.} {\bf 117} (2016) no.~14, 141801},
  \href{http://arxiv.org/abs/1602.01086}{{\tt arXiv:1602.01086 [hep-ph]}}.

\bibitem{Bonaldi:2003ah}
M.~Bonaldi, M.~Cerdonio, L.~Conti, M.~Pinard, G.~A. Prodi, and J.~P. Zendri,
  ``{Selective readout and back action reduction for wideband acoustic
  gravitational wave detectors},''
  \href{http://dx.doi.org/10.1103/PhysRevD.68.102004}{{\em Phys. Rev. D} {\bf
  68} (2003)  102004}, \href{http://arxiv.org/abs/gr-qc/0302012}{{\tt
  arXiv:gr-qc/0302012}}.

\bibitem{PhysRevLett.100.227006}
S.~Sendelbach, D.~Hover, A.~Kittel, M.~M\"uck, J.~M. Martinis, and
  R.~McDermott, ``Magnetism in squids at millikelvin temperatures,''
  \href{http://dx.doi.org/10.1103/PhysRevLett.100.227006}{{\em Phys. Rev.
  Lett.} {\bf 100} (2008) no.~22, 227006}.

\bibitem{kumar2016origin}
P.~Kumar, S.~Sendelbach, M.~A. Beck, J.~W. Freeland, Z.~Wang, H.~Wang, C.~C.
  Yu, R.~Q. Wu, D.~P. Pappas, and R.~McDermott, ``Origin and reduction of $1/f$
  magnetic flux noise in superconducting devices,''
  \href{http://dx.doi.org/10.1103/PhysRevApplied.6.041001}{{\em Phys. Rev.
  Appl.} {\bf 6} (2016)  041001}.

\bibitem{rower2023evolution}
D.~A. Rower, L.~Ateshian, L.~H. Li, M.~Hays, D.~Bluvstein, L.~Ding, B.~Kannan,
  A.~Almanakly, J.~Braum{\"u}ller, D.~K. Kim, {\em et al.}, ``Evolution of
  $1/f$ flux noise in superconducting qubits with weak magnetic fields,''
  \href{http://dx.doi.org/10.1103/PhysRevLett.130.220602}{{\em Phys. Rev.
  Lett.} {\bf 130} (2023)  220602}.

\bibitem{Kubo:1966fyg}
R.~Kubo, ``{The fluctuation-dissipation theorem},''
  \href{http://dx.doi.org/10.1088/0034-4885/29/1/306}{{\em Rept. Prog. Phys.}
  {\bf 29} (1966) no.~1, 255}.

\bibitem{Saulson:1990jc}
P.~R. Saulson, ``{Thermal noise in mechanical experiments},''
  \href{http://dx.doi.org/10.1103/PhysRevD.42.2437}{{\em Phys. Rev. D} {\bf 42}
  (1990)  2437--2445}.

\bibitem{Foster:2017hbq}
J.~W. Foster, N.~L. Rodd, and B.~R. Safdi, ``{Revealing the Dark Matter Halo
  with Axion Direct Detection},''
  \href{http://dx.doi.org/10.1103/PhysRevD.97.123006}{{\em Phys. Rev. D} {\bf
  97} (2018) no.~12, 123006}, \href{http://arxiv.org/abs/1711.10489}{{\tt
  arXiv:1711.10489 [astro-ph.CO]}}.

\bibitem{Benabou:2022qpv}
J.~N. Benabou, J.~W. Foster, Y.~Kahn, B.~R. Safdi, and C.~P. Salemi,
  ``{Lumped-element axion dark matter detection beyond the magnetoquasistatic
  limit},'' \href{http://dx.doi.org/10.1103/PhysRevD.108.035009}{{\em Phys.
  Rev. D} {\bf 108} (2023) no.~3, 035009},
  \href{http://arxiv.org/abs/2211.00008}{{\tt arXiv:2211.00008 [hep-ph]}}.

\bibitem{CONSTABLE2023107090}
C.~Constable and S.~Constable, ``A grand spectrum of the geomagnetic field,''
  \href{http://dx.doi.org/https://doi.org/10.1016/j.pepi.2023.107090}{{\em
  Physics of the Earth and Planetary Interiors} {\bf 344} (2023)  107090}.
  \url{https://www.sciencedirect.com/science/article/pii/S0031920123001164}.

\bibitem{celozzi2023electromagnetic}
S.~Celozzi, R.~Araneo, P.~Burghignoli, and G.~Lovat, {\em Electromagnetic
  shielding: theory and applications}.
\newblock John Wiley \& Sons, 2023.

\bibitem{PhysRevD.75.022002}
L.~Gottardi, ``Complete model of a spherical gravitational wave detector with
  capacitive transducers: Calibration and sensitivity optimization,''
  \href{http://dx.doi.org/10.1103/PhysRevD.75.022002}{{\em Phys. Rev. D} {\bf
  75} (2007) no.~22, 022002}.

\bibitem{Saulson:2017jlf}
P.~R. Saulson, \href{http://dx.doi.org/10.1142/10116}{{\em {Fundamentals of
  Interferometric Gravitational Wave Detectors}}}.
\newblock World Scientific, 2nd. ed.~ed., 2017.

\bibitem{Saulson:1984yg}
P.~R. Saulson, ``{TERRESTRIAL GRAVITATIONAL NOISE ON A GRAVITATIONAL WAVE
  ANTENNA},'' \href{http://dx.doi.org/10.1103/PhysRevD.30.732}{{\em Phys. Rev.
  D} {\bf 30} (1984)  732}.

\bibitem{Hughes:1998pe}
S.~A. Hughes and K.~S. Thorne, ``{Seismic gravity gradient noise in
  interferometric gravitational wave detectors},''
  \href{http://dx.doi.org/10.1103/PhysRevD.58.122002}{{\em Phys. Rev. D} {\bf
  58} (1998)  122002}, \href{http://arxiv.org/abs/gr-qc/9806018}{{\tt
  arXiv:gr-qc/9806018}}.

\bibitem{Grenet:2021vbb}
T.~Grenet, R.~Ballou, Q.~Basto, K.~Martineau, P.~Perrier, P.~Pugnat,
  J.~Quevillon, N.~Roch, and C.~Smith, ``{The Grenoble Axion Haloscope platform
  (GrAHal): development plan and first results},''
  \href{http://arxiv.org/abs/2110.14406}{{\tt arXiv:2110.14406 [hep-ex]}}.

\bibitem{DMRadio:2022pkf}
{\bf DMRadio} Collaboration, L.~Brouwer {\em et al.}, ``{Projected sensitivity
  of DMRadio-m3: A search for the QCD axion below 1\,\,\ensuremath{\mu}eV},''
  \href{http://dx.doi.org/10.1103/PhysRevD.106.103008}{{\em Phys. Rev. D} {\bf
  106} (2022) no.~10, 103008}, \href{http://arxiv.org/abs/2204.13781}{{\tt
  arXiv:2204.13781 [hep-ex]}}.

\bibitem{DMRadio:2023igr}
{\bf DMRadio} Collaboration, A.~AlShirawi {\em et al.}, ``{Electromagnetic
  modeling and science reach of DMRadio-m$^3$},''
  \href{http://arxiv.org/abs/2302.14084}{{\tt arXiv:2302.14084 [hep-ex]}}.

\bibitem{Ouellet:2018beu}
J.~L. Ouellet {\em et al.}, ``{First Results from ABRACADABRA-10 cm: A Search
  for Sub-$\mu$eV Axion Dark Matter},''
  \href{http://dx.doi.org/10.1103/PhysRevLett.122.121802}{{\em Phys. Rev.
  Lett.} {\bf 122} (2019) no.~12, 121802},
  \href{http://arxiv.org/abs/1810.12257}{{\tt arXiv:1810.12257 [hep-ex]}}.

\bibitem{Ouellet:2019tlz}
J.~L. Ouellet {\em et al.}, ``{Design and implementation of the ABRACADABRA-10
  cm axion dark matter search},''
  \href{http://dx.doi.org/10.1103/PhysRevD.99.052012}{{\em Phys. Rev. D} {\bf
  99} (2019) no.~5, 052012}, \href{http://arxiv.org/abs/1901.10652}{{\tt
  arXiv:1901.10652 [physics.ins-det]}}.

\bibitem{Salemi:2021gck}
C.~P. Salemi {\em et al.}, ``{Search for Low-Mass Axion Dark Matter with
  ABRACADABRA-10~cm},''
  \href{http://dx.doi.org/10.1103/PhysRevLett.127.081801}{{\em Phys. Rev.
  Lett.} {\bf 127} (2021) no.~8, 081801},
  \href{http://arxiv.org/abs/2102.06722}{{\tt arXiv:2102.06722 [hep-ex]}}.

\bibitem{Gramolin:2020ict}
A.~V. Gramolin, D.~Aybas, D.~Johnson, J.~Adam, and A.~O. Sushkov, ``{Search for
  axion-like dark matter with ferromagnets},''
  \href{http://dx.doi.org/10.1038/s41567-020-1006-6}{{\em Nature Phys.} {\bf
  17} (2021) no.~1, 79}, \href{http://arxiv.org/abs/2003.03348}{{\tt
  arXiv:2003.03348 [hep-ex]}}.

\bibitem{Devlin:2021fpq}
J.~A. Devlin {\em et al.}, ``{Constraints on the Coupling between Axionlike
  Dark Matter and Photons Using an Antiproton Superconducting Tuned Detection
  Circuit in a Cryogenic Penning Trap},''
  \href{http://dx.doi.org/10.1103/PhysRevLett.126.041301}{{\em Phys. Rev.
  Lett.} {\bf 126} (2021) no.~4, 041301},
  \href{http://arxiv.org/abs/2101.11290}{{\tt arXiv:2101.11290 [astro-ph.CO]}}.

\bibitem{Zhang:2021bpa}
Z.~Zhang, D.~Horns, and O.~Ghosh, ``{Search for dark matter with an LC
  circuit},'' \href{http://dx.doi.org/10.1103/PhysRevD.106.023003}{{\em Phys.
  Rev. D} {\bf 106} (2022) no.~2, 023003},
  \href{http://arxiv.org/abs/2111.04541}{{\tt arXiv:2111.04541 [hep-ex]}}.

\bibitem{Crisosto:2019fcj}
N.~Crisosto, P.~Sikivie, N.~S. Sullivan, D.~B. Tanner, J.~Yang, and G.~Rybka,
  ``{ADMX SLIC: Results from a Superconducting $LC$ Circuit Investigating Cold
  Axions},'' \href{http://dx.doi.org/10.1103/PhysRevLett.124.241101}{{\em Phys.
  Rev. Lett.} {\bf 124} (2020) no.~24, 241101},
  \href{http://arxiv.org/abs/1911.05772}{{\tt arXiv:1911.05772 [astro-ph.CO]}}.

\bibitem{Domcke:2023bat}
V.~Domcke, C.~Garcia-Cely, S.~M. Lee, and N.~L. Rodd, ``{Symmetries and
  selection rules: optimising axion haloscopes for Gravitational Wave
  searches},'' \href{http://dx.doi.org/10.1007/JHEP03(2024)128}{{\em JHEP} {\bf
  03} (2024)  128}, \href{http://arxiv.org/abs/2306.03125}{{\tt
  arXiv:2306.03125 [hep-ph]}}.

\bibitem{Ratzinger:2024spd}
W.~Ratzinger, S.~Schenk, and P.~Schwaller, ``{A Coordinate-Independent
  Formalism for Detecting High-Frequency Gravitational Waves},''
  \href{http://arxiv.org/abs/2404.08572}{{\tt arXiv:2404.08572 [gr-qc]}}.

\end{thebibliography}\endgroup

\clearpage
\newpage
\maketitle
\onecolumngrid
\begin{center}
\textbf{\large Magnets are Weber Bar Gravitational Wave Detectors} \\ 
\vspace{0.05in}
{ \it \large Supplemental Material}\\ 
\vspace{0.05in}
{}
{Valerie Domcke, Sebastian A. R. Ellis, and Nicholas L. Rodd}

\end{center}
\setcounter{equation}{0}
\setcounter{figure}{0}
\setcounter{table}{0}
\setcounter{section}{0}
\renewcommand{\theequation}{S\arabic{equation}}
\renewcommand{\thefigure}{S\arabic{figure}}
\renewcommand{\thetable}{S\arabic{table}}
\renewcommand*{\thesection}{S.\Roman{section}}
\interfootnotelinepenalty=10000 

\setstretch{1.1}

In this Supplemental Material, we derive the experimental sensitivity of our proposed approach in detail.
We begin by computing the mechanical eigenmodes of a coaxial cylinder, which is relevant for understanding the detailed mechanical response of a solenoidal magnet to a passing Gravitational Wave (GW).
From there, we discuss how this effect translates into a signal power spectral density (PSD), and give detailed expressions for the noise PSDs that are used in the main body.
We close with a comparison of our proposal to more traditional mechanical and electromagnetic GW detectors.

Before beginning, we note the conventions adopted throughout this work.
For the GW in the TT frame propagating along the $x$-axis, we fix conventions as
\be
h_{ij}^{\TT} = (h_+ e_{ij}^+  + h_\times e_{ij}^\times ) e^{-i\w(t-x)},
\quad e_{ij}^+ = \frac{1}{\sqrt{2}} 
\begin{pmatrix} 
0 & 0 & 0 \\
0 & 1 & 0 \\
0 & 0 & -1
\end{pmatrix}\!, 
\quad e_{ij}^\times = \frac{1}{\sqrt{2}} 
\begin{pmatrix} 
0 & 0 & 0 \\
0 & 0 & 1 \\
0 & 1 & 0
\end{pmatrix}\!.
\ee
The equivalent result for a GW propagating in other directions can be determined from the appropriate transformations applied to $e_{ij}^{+,\times}$.
To obtain the real-valued gravitational wave the complex conjugate should be added to all expressions linear in the GW, which we drop throughout to avoid cluttered notation.
Additionally, we make use of two-sided PSDs throughout this work, with all PSDs satisfying $S(-f)=S(f)$, implying that the total power in field results after integration of the PSD over $f \in (-\infty,\infty)$.

\section{Interaction of a Gravitational Wave with a Coaxial Cylinder}
\label{supp:MechModes}

In this section we provide the full details of our treatment of the mechanical interaction of a GW with a coaxial cylinder.
We first compute the mechanical eignenmodes for this geometry, then use these to determine the GW overlap, and finally justify Eq.~\eqref{eq:B_master} used in the main text.

\subsection{Mechanical eigenmodes}

In this section we determine the approximate form of the mechanical eigenmodes of a coaxial cylinder.
We take the cylinder to have length $\ell$ as well as inner and outer radii $r_1$ and $r_2$.
This is how the current carrying spool for a solenoid was modelled in the main text and therefore this determines the form of the modes a GW can excite.
In general, if a solid body is defined by a series of positions $\br$, we imagine that after the application of an external force density $\bfo^\text{ext}$, the positions are shifted by $\bdr(\br)$.
This defines a displacement field $\bu(\br) =  \bdr(\br)$, so that $\bu$ is exactly $ \bdr$, although we use this alternative notation as it is common in discussions of mechanical eigenmodes.
Assuming the displacements are parametrically small, then Newton's second law fixes the relation between $\bu$ and $\bfo^\text{ext}$ as
\be
\rho \partial_t^2 \bu = (\lambda + \mu) \nabla (\nabla \cdot \bu) + \mu \nabla^2 \bu + \bfo^\text{ext},
\label{eq:eom_mech_master}
\ee
where $\rho$ denotes the density of the material whereas $\lambda$ and $\mu$ are its Lam\'e coefficients, material quantities which can be related to the Young's modulus and Poisson ratio.
Taking $\bfo^\text{ext}=0$, the above equation determines the mechanical eigenmodes of the system, $\bu_{mnp}$, with $(m,n,p)$ the angular, radial and longitudinal index of the mode.
Each eigenmode is associated with an eigenfrequency $\w_{mnp}$.
Although $\bu$ has dimensions of length, we take $\bu_{mnp}$ to be dimensionless and normalised through
\be
\int_V d^3\br\,\bu_{mnp} \cdot \bu_{m'n'p'}^* = V \delta_{mm'} \delta_{nn'} \delta_{pp'},
\label{eq:norm}
\ee
where the integral is taken over the volume of the coaxial cylinder.
Once the eigenmodes are known, the general solution to Eq.~\eqref{eq:eom_mech_master} in the frequency domain is given by,
\be
\bu(\w) = \sum_{mnp} \left[ \frac{\int d^3\br\, \bfo^\text{ext}(\w) \cdot \bu_{mnp}^*}{\rho V(\w_{mnp}^2-\w^2 - i \w \w_{mnp}/Q_{mnp})} \right] \bu_{mnp}.
\label{eq:bu-general}
\ee
We have regulated the divergence at the resonant frequencies by including finite dissipation, controlled by the mechanical quality factor for each mode, denoted $Q_{mnp}$.

As a first step to determining the eigenmodes of Eq.~\eqref{eq:eom_mech_master}, we perform a Helmholtz decomposition of the displacement field, $\bu_{mnp} = \bu_{L,mnp} + \bu_{T,mnp}$ with $\nabla \times \bu_{L,mnp} = \nabla \cdot \bu_{T,mnp} = 0$.
We can rewrite these fields in terms of a dilatational potential, $\bu_{L,mnp} = \nabla \varphi_{L,mnp}$, and a transverse vector potential, $\bu_{T,mnp} = \nabla \times \bv_{T,mnp}$.
If $\rho$ is constant, then Eq.~\eqref{eq:eom_mech_master} can be written as a pair of Helmholtz equations for the potentials $\varphi_{L,mnp}$ and $\bv_{T,mnp}$.
For a coaxial cylinder the general solution to these equations can be written as,
\begin{align}
(\varphi_L)_{mnp} & = f_L(r) \cdot [e \cos(m \phi) + (1 - e) \sin(m \phi)] \cdot \left[ a \cos\left(\frac{p \pi z}{\ell}\right) + (1 - a) \sin\left(\frac{p \pi z}{\ell}\right)\right]\!, \\
(\bv_T)_{mnp} & = f_r(r) \cdot [e_r \cos(m \phi) + (1 - e_r) \sin(m \phi)] \cdot \left[ a_r \cos\left(\frac{p \pi z}{\ell}\right) + (1 - a_r) \sin\left(\frac{p \pi z}{\ell}\right) \right] \he_r \nonumber \\
& -   f_r(r) \cdot [(1 - e_r) \cos(m \phi) + e_r \sin(m \phi)] \cdot (1 - 2e_r) \left[a_r \cos\left(\frac{p \pi z}{\ell}\right) + (1 - a_r) \sin\left(\frac{p \pi z}{\ell}\right) \right] \he_\phi  \nonumber \\
& + f_z(r) \cdot [e_z \cos(m \phi) + (1 - e_z) \sin(m \phi)] \cdot \left[ a_z \cos\left(\frac{p \pi z}{\ell}\right) + (1 - a_z) \sin\left(\frac{p \pi z}{\ell}\right) \right] \he_z,
\end{align}
where the radial functions are specified by
\begin{align}
f_L(r) &= c_o J_m(q_{m n} r) + c_i Y_m(q_{mn} r), \nonumber \\ 
f_r(r) &= c_{or} J_{m+1}(k_{mn} r) + c_{ir} Y_{m+1}(k_{mn} r), \\
f_z(r) &= c_{oz} J_{m}(k_{mn} r) + c_{iz} Y_{m}(k_{mn} r). \nonumber
\end{align}
Here $(r, \phi, z)$ indicate the position within the spool in cylindrical coordinates, $J_m$ and $Y_m$ denote Bessel functions, and $e, e_r, e_z, a, a_r, a_z \in \{0, 1 \}$ select the respective odd and even components of the mode.
The values of $q_{mn}$ and $k_{mn}$ determine the eigenfrequencies and are computed below.
The coefficients $c_o, c_{or}, c_{oz}, c_i, c_{ir}, c_{iz}$ in the radial functions are set by boundary conditions and the overall normalisation convention, see below.
The correlation of the radial and azimuthal components of the transverse vector potential is a consequence of the cylindrical symmetry of the problem.

By construction, the eigenmodes are solutions to the homogeneous equation, and therefore the net force on the system must vanish for these modes.
In detail,
\be
\int dV\,f_i^\text{ext} = \int_{\partial V} dS\, n_k \sigma_{ik} = 0,
\label{eq:boundary}
\ee
where $\partial V$ is the surface of the material (in our case a coaxial cylinder) and $\mathbf{n}$ denotes the normal vector to this surface.
We have further introduced the stress tensor $\sigma_{ij}$, which can be written in terms of the strain tensor $u_{ij}$ as follows,
\be
\sigma_{ij} = \lambda u_{kk} \delta_{ij} + 2 \mu u_{ij}, 
\quad 
u_{ij} \equiv \frac{1}{2} \left( \frac{\partial u_i}{\partial x_j} + \frac{\partial u_j}{\partial x_i} \right)\!.
\ee
Due to the symmetries of the trigonometric functions in the expressions above, Eq.~\eqref{eq:boundary} alone is not in general sufficient to fully determine the free coefficients for a given mode.
To identify one possible solution among this family of solutions, we impose a stronger, local boundary condition
\begin{align}
n_k \sigma_{ik}(\br_0) = 0,
\label{eq:boundary_local}
\end{align}
for some points $\br_0$ on the boundary $\partial V$.
For our purpose, this subset of mode functions provide representative examples that are sufficient for understanding the general properties of the modes.

\begin{figure}[t]
\centering
\includegraphics[width = 0.4 \textwidth]{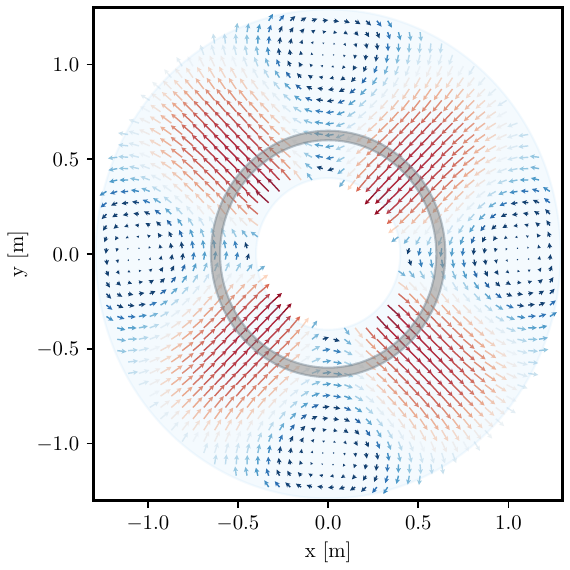} \hspace{5mm}
 \includegraphics[width = 0.4 \textwidth]{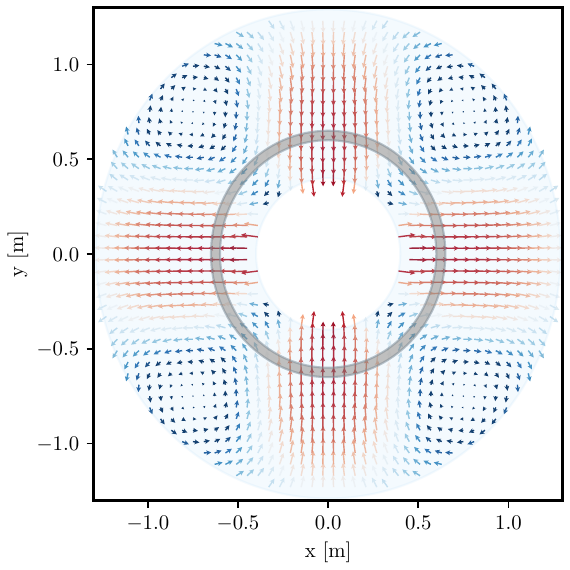}
\caption{The $mnp=210$ eigenmode of a coaxial cylinder for even ($e_z = 1$, left) and odd ($e_z = 0$, right) angular dependence.
The grey shaded area indicates the cylinder cross section we consider, although we show the modes over a larger range to highlight their structure.
The close resemblance with the GW $\times$ (left) and $+$ (right) polarisations for a GW incoming along the $z$-axis is indicative of a high overlap factor in this case.
In both plots the color scheme is such that red corresponds to larger values of the displacement and blue smaller.
See text for the details of how these modes are computed.}
\label{fig:210-mode}
\end{figure}

As an example, consider the set of modes with $mnp = 2n0$.
Setting $a = a_r = 0$ trivially satisfies Eq.~\eqref{eq:boundary_local} on both end caps of the cylinder, $z = \pm \ell/2$.
This greatly simplifies the potentials, 
\be
\varphi_L  = 0,
\quad
\bv_T  = f_z(r) \cdot [e_z \cos(2 \phi) + (1 - e_z) \sin(2 \phi)] \he_z,
\ee
where we have taken $a_z=1$ to avoid the trivial solution.
To determine the coefficients in $f_z(r)$, we impose the local boundary condition at a radial boundary $r_0$, which enforces
\be
F_r(r_0) [e_z \sin(2 \phi) - (1-e_z) \cos(2\phi)] \he_r - F_\phi(r_0) [e_z \cos(2 \phi) + (1-e_z) \sin(2\phi)] \he_\phi = 0,
\label{eq:radial_boundary}
\ee
with
\begin{equation}\begin{aligned}
F_r(r) & = - 4 \{c_{oz} [x J_1(x) - 3  J_2(x)] + c_{iz}[x Y_1(x) - 3 Y_2(x)]\}, \\
F_\phi(r) &= \frac{1}{x} \{c_{oz} [(x^2 - 12) x J_0(x) - 4 (x^2 - 6) J_1(x)] + c_{iz} [(x^2 - 12) x Y_0(x) - 4 (x^2 - 6) Y_1(x)]\}.
\end{aligned}\end{equation}
where $x = k_{2n} r$.
Equation~\eqref{eq:radial_boundary} cannot be fulfilled at the inner radial boundary $r_0 = r_1$ and the outer radial boundary $r_0 = r_2$ for non-trivial $c_{oz}$ and $c_{iz}$ for all values of $\phi$. 
However, since Eq.~\eqref{eq:boundary} is ensured to be fulfilled, we are free to pick a particular value of $\phi$ to impose the stronger condition Eq.~\eqref{eq:boundary_local} or equivalently Eq.~\eqref{eq:radial_boundary}.
For the even $e_z=1$ and odd $e_z=0$ modes, we take $\phi = \pi/4$ and $0$, so that the condition to be satisfied is $F_r(r_1) = F_r(r_2) = 0$.
Non-trivial solutions $c_{oz,iz}$ to this system of equations can be obtained if the corresponding determinant vanishes.
Since the determinant is a combination of Bessel functions with arguments $k_{2n} r_1$ and $k_{2n} r_2$, the $n$th zero of this determinant fixes the eigenfrequency of the mode through the relation\footnote{Varying the choice of $\phi$ does impact the modes, although we have confirmed the change to the numerical value of the eigenfrequency and the shape of the $n=1$ modes is small.}
\be
\w_{mnp}^2 = (k_{mn}^2 + (p \pi/\ell)^2 ) \mu/\rho.
\ee
For $r_1 = 0.6$~m, $r_2 = 0.65$~m and picking the lowest radial mode $n = 1$ this results in $c_{iz} = - 0.072 \, c_{oz}$.
Again, the modes are normalised with Eq.~\eqref{eq:norm}.
In Fig.~\ref{fig:210-mode} we visualise the resulting eigenmodes in the $x$-$y$-plane, in this case there is no displacement in the $z$-direction. 
For stainless steel ($\rho = 7.96$~g/cm$^3$, $\lambda = 90.8$~GPa, $\mu = 77.4$~GPa), the eigenfrequency of this mode is found to be 8.6~kHz, roughly corresponding to a value of $k_{21}$ given by the diameter of the cylinder.
For better visualisation, we show displacement vectors over a large radial volume, though we note that in our computation here we only consider the eigenmodes of the narrow gray region indicating the position of the conducting wires.
In a less idealised setup, the motion of the entire support structure and its coupling to the conducting wires should be taken into account.

\subsection{Overlap factors}

\begin{figure}[t]
\centering
\includegraphics[width = 0.4 \textwidth]{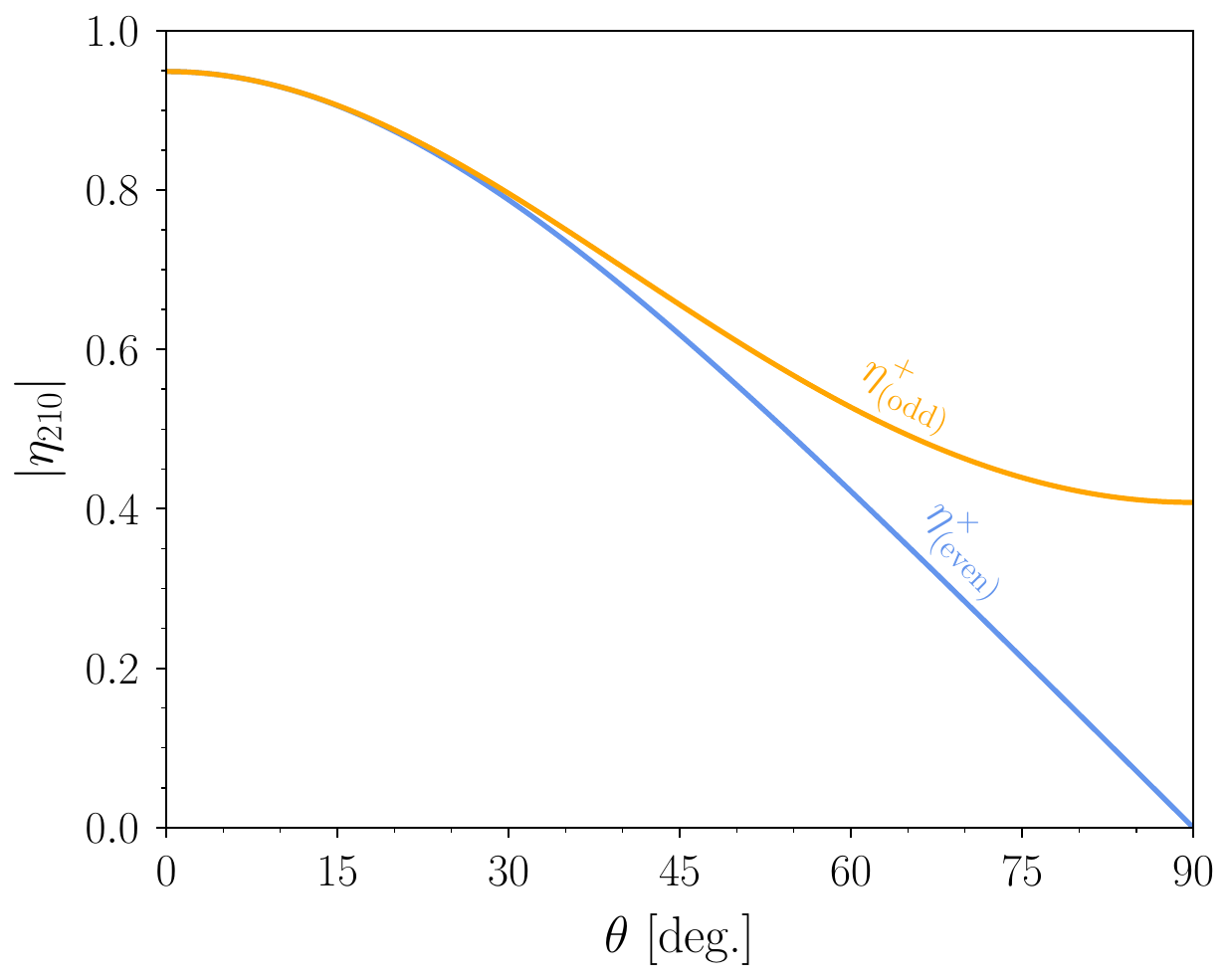}
\caption{The overlap factors between the a GW incoming at an angle $\theta$ from the symmetry axis of a cylinder and the even ($e_z = 1$, blue) and odd ($e_z = 0$, orange) eigenmode.
The $+$ polarisation of the GW couples only to the odd 210 mode, while the $\times$ polarisation couples only to the even 210 mode.}
\label{fig:overlap210}
\end{figure}

As outlined above, once the eigenmodes are known, the mechanical response of the cylinder to an external force can be determined from Eq.~\eqref{eq:bu-general}.
The specific force of interest is that induced by a GW in the proper detector frame, which generates a force density on the system of $f_i = \tfrac{1}{2} \rho \ddot{h}_{ij}^{\TT} x^j$.
Substituting this in, we have
\be
\bu = h_A \sum_{mnp} \left[ \frac{\w^2 \ell_{\eta} \eta^A_{mnp}}{\w^2-\w_{mnp}^2 + i \w \w_{mnp}/Q_{mnp}} \right] \bu_{mnp},
\label{eq:bu-GW-sum}
\ee
where $A=+,\times$ is summed over and, as in the main text, we have defined a dimensionless overlap factor
\be
\eta_{mnp}^A = \frac{1}{2V \ell_{\eta}} \int_V d^3 \br\, e_{ij}^A \br^i[\bu_{mnp}^*]^j,
\ee
with the normalisation factor $\ell_{\eta}^2 = \{\ell^2 + 9(r_1^2 + r_2^2) + [3(r_1^2 + r_2^2) - \ell^2]\cos(2\theta)\}/192$ chosen such that if we replaced the eigenmode in the above integral by the spatial structure of the GW force, $[\bu_{mnp}^*]^j \to \tfrac{1}{2} e_{ij}^A \br^j/\ell_\eta$, then we would obtain $\eta = 1$.
Here $\theta$ represents the angle the GW is incident with respect to the $z$-axis of the cylinder.
As defined, an overlap factor smaller (larger) than unity thus indicates a coupling which is less (more) efficient than in the free-falling limit.
The numeric overlap factors for the 210 mode are shown in Fig.~\ref{fig:overlap210} for $e_z=1$ (even) and $e_z=0$ (odd)
We see that for a GW incident along the $z$-axis, the overlap factor reaches $\eta^{+,\times}_{210}(\theta=0) \simeq 0.95$, indicating a very efficient coupling between the GW and the eigenmode.

\subsection{Induced magnetic field}
\label{subsec:IMF}

In the magnetostatic limit $\w \ll \{\ell^{-1}, r_1^{-1}\}$, the magnetic field induced by single loop ${\cal C}$ carrying a current $I$ is determined by the Biot-Savart law as,
\be
\bB(\br') = \frac{I}{4 \pi} \int_{\cal C} \frac{r \, d\boldsymbol{{\cal C}} \times (\br' - \bxi)}{|\br' - \bxi|^3},
\ee
with $d\boldsymbol{{\cal C}}$ the infinitesimal line element along the loop
\be
d\boldsymbol{{\cal C}} = \frac{d \bxi}{d \phi} \left| \frac{d \bxi}{d \phi} \right|^{-1}  d\phi.
\ee
As in the main text, $\bxi = \br +  \bdr$ parametrises the position of the loop and $\br'$ the position where we evaluate the magnetic field, which we envision as a location within the pickup loop.
In the proper detector frame, the GW imparts a Newtonian force on the loop.
This force changes the position of the loop and hence its orientation, with the latter encoded in $d \boldsymbol{{\cal C}}$ (as the GW displacement is small, we can continue to parameterise the curve with $\phi$).
In the absence of the GW, $\bxi = \br$ and $d\boldsymbol{{\cal C}} = \he_{\phi} d\phi$, recovering the flat space Biot-Savart result.

The above calculation can be immediately extended to a spool with $N$ windings, spanning a finite volume in the axial and radial direction.
Doing so yields
\be
\bB(\br') = \frac{I}{4 \pi} \frac{N_z}{\ell} \frac{N_r}{\Delta r} \int_{r_1}^{r_2} dr\,\int_{- \ell/2}^{\ell/2} dz\,  \int_{\cal C} \frac{ r \, d\boldsymbol{{\cal C}} \times (\br' - \bxi)}{|\br' - \bxi|^3},
\label{eq:B_app}
\ee
where $\Delta r = r_2-r_1$ and the number of windings in the axial and radial direction is denoted $N_{z,r}$, with the total current per winding $I$ unchanged by a passing GW.
Identifying $I_N = N_z N_r I$ yields the expression given in Eq.~\eqref{eq:B_master} of the main text.

For convenience, we give the explicit expressions for the Cartesian components of the integrand ${\cal I}_{x,y,z}^{+, \times}(r, \phi, z)$ in Eq.~\eqref{eq:B_app} in the free falling limit for a cylinder aligned along the $z$-axis with a GW incoming along the $x$-axis, where the superscript distinguishes the response to the two GW polarisations,
\begin{equation}\begin{aligned}
(2 \sqrt{2} \rho^5) {\cal I}_x^\times & = 6 r (z_0' - z)^2 c_\phi (r s_\phi - r_0' s_{\phi_0'}), \\
(2 \sqrt{2} \rho^5) {\cal I}_y^\times & = - r [(r - r_0' c_{\phi - \phi_0'})\rho^2 - 6 (z_0' - z)^2 s_\phi (r s_\phi - r_0' s_{\phi_0'})], \\
(2 \sqrt{2} \rho^5) {\cal I}_z^\times & =  r (z_0' - z) [( 5 r^2 + 2 r_0'^2 - (z_0' - z)^2) s_\phi - r_0' (3 r_0' s_{\phi - 2 \phi_0'} + 2 r (s_{2 \phi - \phi_0'} + 4 s_{\phi_0'}))],
\end{aligned}\end{equation}
and
\begin{equation}\begin{aligned}
(2 \sqrt{2} \rho^5) {\cal I}_x^+ & = - r (z_0' - z) c_\phi \{\rho^2 c_\phi^2 - 3[(z_0' - z) + r s_\phi - r_0' s_{\phi_0'}][(z_0' - z) - r s_\phi + r_0' s_{\phi_0'}]\}, \\
(2 \sqrt{2} \rho^5) {\cal I}_y^+ & = - \tfrac{1}{4} r (z_0' - z) \{\rho^2 (5 s_\phi + s_{3 \phi}) - 12 s_\phi[(z_0' - z) + r s_\phi - r_0' s_{\phi_0'}][(z_0' - z) - r s_\phi + r_0' s_{\phi_0'}]\}, \\
(2 \sqrt{2} \rho^5)  {\cal I}_z^+ & = - r(r-r_0'c_{\phi - \phi_0'})[- 3 (z_0' - z)^2 + (3 r^2 - \rho^2) s_\phi^2 - 6 r r_0' s_\phi s_{\phi_0'} + 3 r_0'^2 s_{\phi_0'}^2],
\end{aligned}\end{equation}
with $\rho^2 = r^2 + r_0'^2 + (z_0' -z)^2 - 2 r r_0' \cos(\phi - \phi_0')$ and where we employ the shorthand $c_x = \cos x$ and $s_x = \sin x$.
Here $r_0', \phi_0', z_0'$ denote the flat space coordinates ($\br_0'$ in the main text).
To compute the flux through a pickup loop, one must integrate over the area of the loop as it is perturbed by the GW.
However, it is convenient to transform this integral back to the flat space coordinates where the area of the loop is straightforward.

\begin{figure}[t]
\center
\includegraphics[width = 0.42 \textwidth]{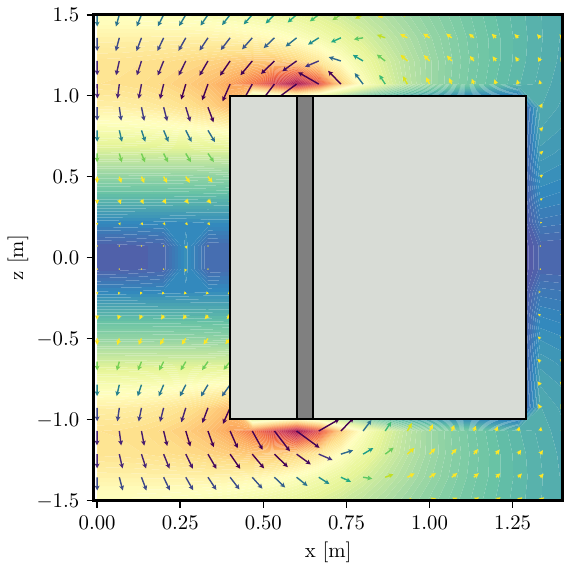} 
\hfil
\includegraphics[width = 0.42 \textwidth]{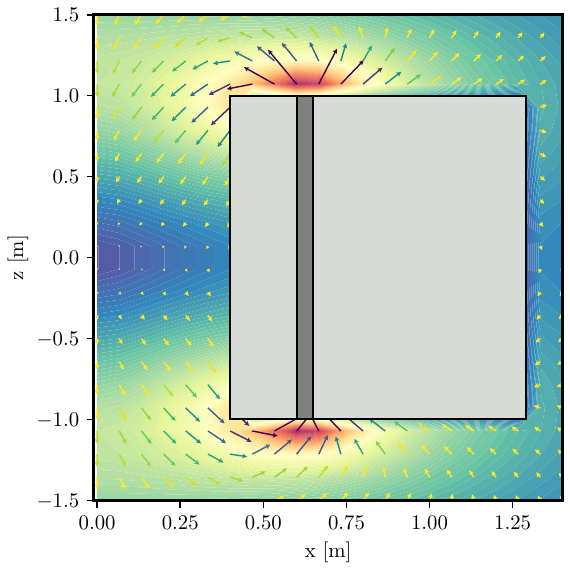}
\caption{Magnetic field in the $x$-$z$-plane induced by a GW incoming along the $x$-axis with $+$ polarisation in the low-frequency approximation (left) and in the resonant regime (right, 210 mode).
Red contours correspond to the highest value of the magnetic field, while blue are the lowest. Arrows indicate the magnitude and direction of the field. The light gray shaded regions indicate the position of the magnet structure, while the dark gray indicates the position of the current-carrying spool.}
\label{fig:Bh}
\end{figure}

\begin{figure}[t]
\center
\includegraphics[width = 0.42 \textwidth]{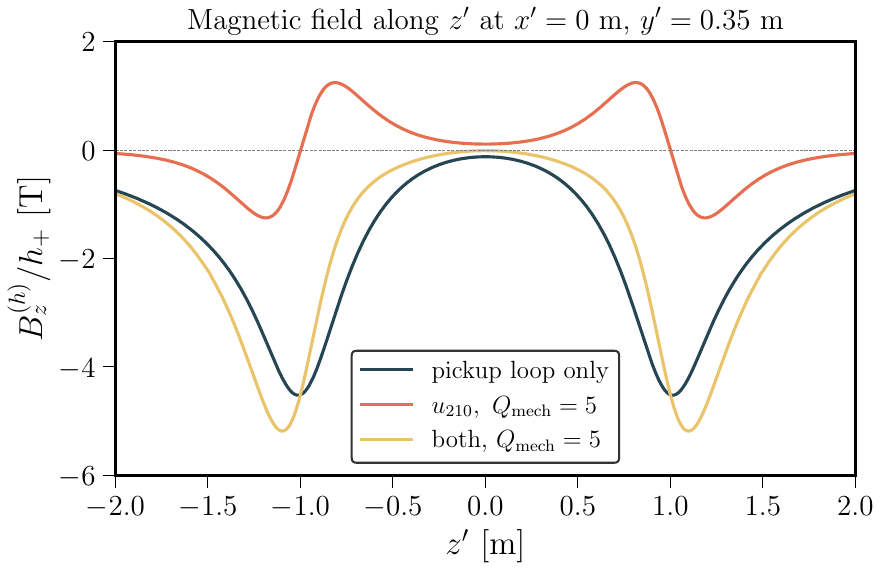} 
\hfil
\includegraphics[width = 0.42 \textwidth]{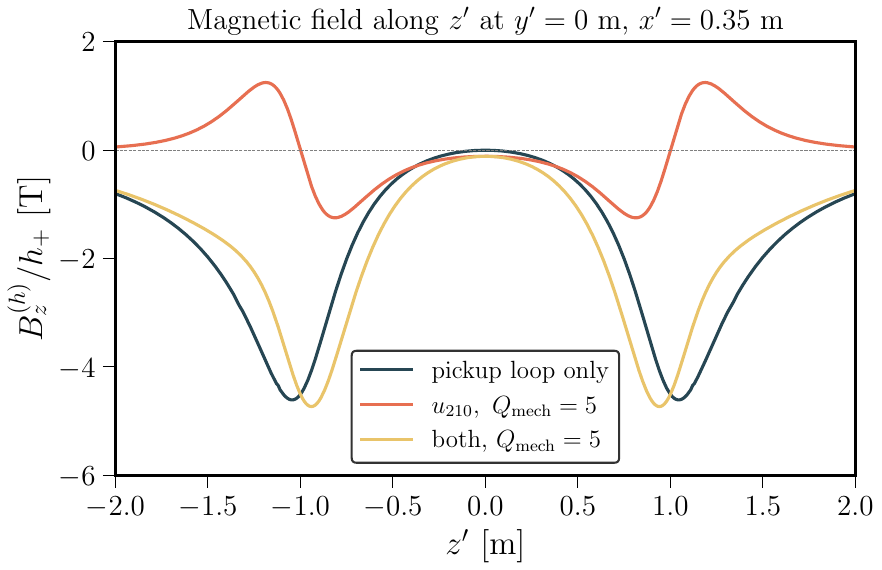}
\caption{Magnetic field along the $z'$ direction at (left) $y' = 0.35\,\text{m},~x'=0\,\text{m}$ and (right) $y' = 0\,\text{m},~x'=0.35\,\text{m}$ induced by a GW incoming along the $x$-axis with $+$ polarisation with flexible pickup loop (black), rigid pickup loop with flexible mechanical mode $u_{210}$ on resonance (red) and both flexible on resonance (gold).
For the mechanical mode contributions we have set $Q_{\rm mech} = 5$ for better visualisation.}
\label{fig:Bhz}
\end{figure}

A cross-section of the resulting induced magnetic field is depicted in Fig.~\ref{fig:Bh} for the magnet dimensions given in the main text.
The induced field along the $z$-axis is shown in Fig.~\ref{fig:Bhz}.
In Fig.~\ref{fig:Bh}, the left panel represents the field seen resulting from a change only in the relative positioning of the pickup loop and magnet, while the right panel represents the field induced by the coupling of the GW to the odd 210 mechanical mode.
The latter serves as an illustration of the effect of GWs on the magnet structure.
(In the right panel the pickup loop remains free falling, however the signal is dominated by the mechanical mode of the magnet by focussing on a mechanical resonance.)
At frequencies below the lowest-lying mechanical resonance, the magnet can be treated as rigid and the effect of the pickup loop positioning is expected to dominate, while at higher frequencies, the full effect is a superposition of mechanical mode responses and the loop position.
In Fig.~\ref{fig:Bhz}, we show the $z$-component of the signal magnetic field due to a $+$-polarised GW arriving along the $x$-axis of the detector.
In black we show the magnetic field in the regime where the magnet is effectively rigid, and only the relative motion of the pickup loop contributes to the signal.
In red we show the induced magnetic field by the coupling of the GW to the $210$ mechanical mode alone, assuming a (unphysical) rigid pickup loop.
In gold we show the effect of the GW coupling to both the pickup loop and the $210$ mechanical mode assuming a mechanical $Q_{\rm mech} = 5$.
Several features are important to note.
The first is that both the contribution from the pickup loop and the contribution from the 210 mode of the magnet are maximal close to the end caps of the magnet at $z' \sim 1$~m. Closer to the center of the magnet the relative motion of pickup loop and magnet coils is suppressed, whereas outside the bore of the magnet the background magnetic field drops off.
The second is that the $210$ mode-induced magnetic field changes sign along the $z'$ direction, leading to partial cancellation with the contribution from the movement of the pickup loop.
The third is that the $210$ mode-induced contribution at a given $z'$ position changes sign depending on whether one evaluates it along the $x'$ or the $y'$ axis.
Indeed, this is to be expected on the basis of the shape of the mode, shown in Fig.~\ref{fig:210-mode}, which exhibits the characteristic quadrupolar structure expected to couple strongly to a GW.
This leads to a full cancellation of the mode contribution if one were to integrate over a circular loop, and motivates us to consider the quarter-circle pickup loop placed close to the end caps of the cylinder below.

\section{Signal Power Spectral Density}

In this section we demonstrate how to derive a simple analytic expression for the power spectral density of either the flux $\Phi_h$ through a pickup loop of area $A$ induced by the GW, or of the electromotive force (EMF) $\mathcal{E}_h \equiv - \partial_t \Phi_h$.
We start from the assumption that the pickup loop is placed normal to the solenoidal axis, i.e. $\hat{\mathbf{n}} = \he_z$.
The plane of the loop will be unaffected by $+$-polarized GWs, but $\times$-polarized waves will tilt the plane. As such, we must consider the three components of the GW-induced magnetic field $\bB_h$, given in Eq.~\eqref{eq:B_app}, which can be re-expressed as
\begin{equation}\begin{aligned}
B_{h,i}(t) &= h_A e^{-i \omega t}\,\left[\frac{B_0}{2\pi \ell \ln F} \int r\,dr\,dz\, \frac{(d\boldsymbol{{\cal C}} \times (\br' - \bxi))\cdot\he_i}{|\br'-\bxi|^3} \bigg|_{{\cal O}(h_A)} \right]\!,
\\
&F \equiv \frac{2r_2 + \sqrt{\ell^2 +4r_2^2 }}{2 r_1 + \sqrt{\ell^2 +4 r_1^2}},
\label{eq:BzhGen}
\end{aligned}\end{equation}
where $B_0$ is the peak flat-space magnetic field at the origin, $B_0 = I_N \, \ln F /2\Delta r $ in the notation of the preceding section.
In the first expression we have introduced the shorthand $\mathcal{O}(h_A)$, meaning that for a given polarisation we expand the expression in square brackets to linear order in $h$ and keep only the coefficient of that term (while setting $t=\bx=0$ for the GW).

The flux and EMF through a pickup loop of flat space area $A_p$ are given by
\be
\Phi_h = \left[ \gamma \int d\mathbf{A}_p\,\cdot \bB_{h} \right] \bigg|_{{\cal O}(h_A)}, 
\hspace{0.5cm}
\gamma = 1 + \frac{1}{2\sqrt{2}} h_+ s_{\theta}^2 e^{-i \omega t},
\hspace{0.5cm}
\mathcal{E}_h = - \partial_t \Phi_h.
\label{eq:Phih-withJac}
\ee
Here $\gamma$ is a Jacobian that results from transforming the area of the loop as perturbed by the GW back to the flat space coordinates we integrate over.
Examining Eqs.~\eqref{eq:BzhGen} and \eqref{eq:Phih-withJac} above, we see that the only time-dependence arises due to the GW, meaning that we can easily compute flux and EMF PSD, obtaining
\begin{equation}\begin{aligned}
&S_\Phi(\w) = B_0^2 A_p^2 \Bigg\vert \Bigg[\frac{\gamma}{2\pi \ell \ln F A_p} \int d\mathbf{A}_p\,\cdot\, \int r\,dr\,dz\,\frac{(d\boldsymbol{{\cal C}} \times (\br' - \bxi))}{|\br'-\bxi|^3} \Bigg] \bigg|_{{\cal O}(h_A)} \Bigg\vert^2 S_h(\w) \equiv  B_0^2 A_{p}^2 |\mathcal{G}|^2 S_h(\w), \\
&S_\mathcal{E}(\w) = \w^2 S_\Phi(\w).
\end{aligned}\end{equation}
The first line defines a dimensionless gain factor $\mathcal{G}$ that and must be computed numerically, and is a position and frequency-dependent quantity.
In our broadband configuration, the flux in the pickup loop is read out through a SQUID magnetometer, and to determine the flux observed by the SQUID we must account for the inductive coupling between the systems, leading to
\begin{align}
S_{\rm sig}(\omega) = \frac{\alpha^2}{4}\frac{L}{L_p}B_0^2 A_{p}^2 |\mathcal{G}|^2 S_h(\w).
\label{eq:sigPSD}
\end{align}
Here $L$ and $L_p$ are the inductances of the SQUID and pickup loop, whereas $\alpha$ is a coupling coefficient between the two systems, with a characteristic values being $L \simeq 1\,\text{nH}$ and $\alpha \sim 1/\sqrt{2}$~\cite{Kahn:2016aff}.
The above indicates we should minimise $L_p$ to maximise $S_{\rm sig}$.
This is only possible up to the minimum value of $L_p$ consistent with energy conservation (as $L_p \to 0$ the energy stored in the inductor diverges).
For a solenoid with a coaxial pickup loop, the minimum inductance is $L_p \sim \pi R_p^2/\ell$, where again $R_p$ is the radius of the pickup loop.
(This minimum value is the same as determined for a toroidal geometry in Ref.~\cite{Kahn:2016aff}, although we note that for different configurations, e.g. if the pickup loop is not coaxial, the minimum inductance varies.)

\begin{figure}[t]
\center
\includegraphics[width = 0.42 \textwidth]{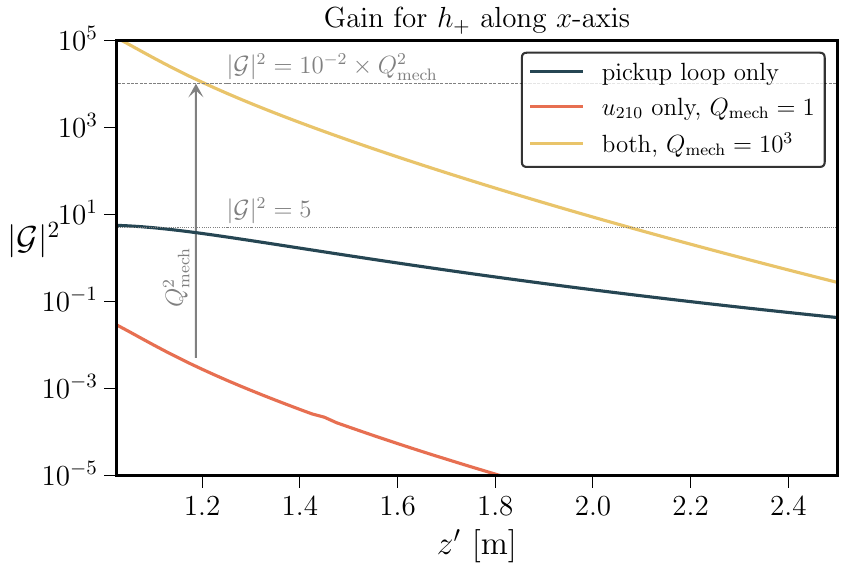}
\caption{Position-dependent gain $|\mathcal{G}|^2$ for three scenarios: rigid magnet with flexible pickup loop (black), rigid pickup loop with flexible mechanical mode $u_{210}$ off-resonance (red), and a flexible loop and mechanical mode on resonance (gold).
We assume a $+$-polarised GW propagating along the $x$-axis.
For the pickup loop only curve, we assume a coaxial circular loop of radius equal to the bore of the ADMX-EFR magnet, $R_p = 0.4$\,m.
For the mode contributions, we assume a loop placed in a quadrant from $\phi' = [-\pi/4, \pi/4]$ from $r' = [0.5\,\text{m}, 0.75\,\text{m}]$ to maximise the gain.
}
\label{fig:GainCoaxialLoop}
\end{figure}

We plot the position dependence of $\mathcal{G}$ in Fig.~\ref{fig:GainCoaxialLoop} both in the low frequency regime (black), where we approximate the magnet to be rigid, and close to a mechanical resonance (gold, $Q_\text{210} = 10^3$), where we model the motion of the magnet by the 210-mode as discussed above.
For comparison, we show in red the magnet response above the resonance frequency ($Q_\text{mech} = 1$) while treating the pickup loop as rigid.
From the right panel of Fig.~\ref{fig:Bh} (see also Fig.~\ref{fig:Bhz}), it is clear that the $z$-component of the induced magnetic field is maximised just outside the bore of the magnet ($z' > 1$), near the position of the current-carrying spool.
Therefore, for illustrating the maximum gain due to the 210 mode, we have assumed a pickup loop spanning one quadrant, from $\phi' \in [- \pi/4,\pi/4]$ placed off-centre such that it covers the radial region $r' = [0.5\,\text{m}, 0.75\,\text{m}]$.
Meanwhile, for the curve illustrating the effect due to the pickup loop only (black), we have chosen a coaxial circular pickup loop. 
In practice, two pickup loops could be placed at opposite ends of the magnet to avoid parasitic mutual inductance, with the two optimised for the off- and on-resonance signals respectively.  Placing the loop outside of the bore of the magnet has the added advantage of being able to consider larger pickup loop radii and allowing the co-use of the magnet with other experiments.

From Fig.~\ref{fig:GainCoaxialLoop}, we make several observations:
(i) The peak gain factor for the loop contribution is close to unity (more precisely $|{\cal G}|_\text{loop,max}^2 \simeq 5$), while the contribution from the 210 magnet mode (off-resonance) is suppressed. 
This is due to the placement of the loop (normal to the $z$-axis) and the particular choice of mode, which features no displacement in $z$-direction.
Consequently, for this particular mode, the dominant effect of `stretching' the cylinder in $z$-direction is absent and instead we are picking up the small changes in the magnet field due to deformation of the cylinder coils normal to the $z$-axis as shown in Fig.~\ref{fig:210-mode}.
In addition, for a GW propagating in the $x$-direction, we have accounted for the suppression factor $(\eta_{210}^+)^2 \simeq 0.17$ from the overlap factor.
Considering a range of modes, or different orientations of the pickup loop, we expect to recover an order one gain factor for the off-resonance mode contribution.
(ii) Due to the symmetry structure of the 210 mode, see Fig.~\ref{fig:210-mode}, the leading order magnet mode contribution vanishes when integrating over a circular coaxially aligned pickup loop.
More generally, a range of modes and GWs originating from different directions will contribute to the signal, requiring a more detailed investigation to determine the optimal choice of shape and location of the pickup loop.
Similar sensitivity to the geometric properties of the GW occurs in classical resonant mass detectors, and can be used to distinguish GWs from other vibrational noise sources~\cite{Bonaldi:2003ah}.
(iii) Importantly, we note that as expected, on resonance the gain factor is enhanced by $|{\cal G}|^2 \propto Q_\text{mech}^2$. Here we have chosen a small quality factor, $Q_\text{mech} = 10^3$ to keep all curves on the same plotting range.
Quality factors of $Q_\text{mech} = 10^6$ and larger have been achieved for large resonant bars~\cite{Vinante:2006uk}, leading to the significant enhancement on resonance indicated Fig.~\ref{fig:ShPlotMagnet}. 
We thereby justify the scaling of the gain used in the main text.

\section{Flux quantization}

We can use the formalism developed above to compute the flux through a superconducting current carrying coil of the magnet in the presence of GW.
Consider a superconducting loop of radius $R$ in the $x-y$-plane carrying a persistent current $I$.
Working in the free-falling regime, as discussed above, the total flux through this loop is given by
\be
\Phi
= \gamma \int d\mathbf{A}_I\, \cdot\,\bB(\br'_0),
\ee
where in contrast to the earlier calculation the integral is performed over the area of the loop that carries the current, $A_I$.
Evaluating this, we find that the integrand vanishes at ${\cal O}(h)$.
This has two important consequences.
First, it ensures that flux through the current carrying loop is conserved.
In other words, both the area of the loop and the magnetic field are modified at ${\cal O}(h)$, but in such a manner that the flux through the current carrying loop remains conserved.
This is critical given the flux conservation condition obeyed by supercurrents given that a GW force is too small to change the flux by integer units of a flux quanta.
Second, it implies that no flux change is induced by a GW through any coplanar coaxial pickup loop placed in the bore of the magnet as long as the magnet and pickup loop are both in the free-falling regime.
As discussed above a flux is however induced once there is relative motion between the magnet and the pickup loop.
This occurs, for instance, with a pickup loop placed outside of the magnet bore, a pickup loop not placed coaxially, or for frequencies at which the magnet and pickup loop response differ.

\section{Noise Power Spectral Densities}

Having determined the signal PSD, we next consider the relevant noise sources in the case of using a broadband or resonant readout of the pickup loop.
We reiterate that for simplicity the magnet we have envisioned using is a persistent superconducting device.
For other magnets a careful consideration of the noise contribution from an external power supply would be required.

\subsection{Broadband setup noise}

The simplest experimental setup would consist of a pickup loop inductively coupled to a SQUID.
In this case, the noise of the SQUID is expected to dominate, with a power spectral density given at frequencies above $f\sim 10\,\text{Hz}$ by~\cite{PhysRevLett.100.227006,kumar2016origin, rower2023evolution}
\be
S_{\Phi}^{\SQ} = 10^{-12} \Phi_0^2 /\text{Hz} \simeq 4.28 \times 10^{-42}\,\text{Wb}^2/\text{Hz},
\label{eq:squidPSD}
\ee
where $\Phi_0 = \pi \hbar / e\simeq 2.07 \times 10^{-15}\,\text{Wb}$ is the flux quantum, with the last equality given in SI units.

Additional noise from thermal vibrations of the magnet and the pickup loop could also be important.
We estimate these as follows.
The fluctuation-dissipation theorem implies the magnet will experience a force PSD of $S_{F}^{\rm th} = 2 M T \w_{\rm mech}/Q_{\rm mech}$ \cite{Kubo:1966fyg}, and therefore an acceleration PSD of $S_{F}^{\rm th} M^{-2}$.
We note that the superconducting nature of the magnet justifies this treatment as a uniform block of material, rather than a collection of loose parts.
Accounting for the frequency response of the magnet, the displacement PSD associated with the thermal vibrations under the assumptions that the forces are independent of position is\footnote{Near the resonance, where the thermal noise is relevant, this approximation is a good one, see Ref.~\cite{Saulson:1990jc}.}
\be
S_x^{\rm th. mech.} \simeq \frac{S_{F}^{\rm th} M^{-2} }{(\w_{\rm mech}^2 - \w^2)^2 + (\w_{\rm mech} \w/Q_{\rm mech})^2}.
\label{eq:mechPSDx}
\ee
The displacements perturb the length of the magnet and therefore for a displacement $x$ we can estimate the size of the induced magnetic field as $\sim B_0 x/\ell$, similar to the simple estimate we performed for the GW in the introduction.
As seen by the SQUID, the resulting vibrational noise source scales as
\be
S_{\Phi}^{\rm th. mech.} \sim \frac{\alpha^2}{4}\frac{L}{L_p}\,\frac{B_0^2 A_{p}^2}{\ell^2} \frac{S_{F}^{\rm th} M^{-2} }{(\w_{\rm mech}^2 - \w^2)^2 + (\w_{\rm mech} \w/Q_{\rm mech})^2}.
\label{eq:mechPSD}
\ee
For an $M= 40$\,ton bar of length $\ell = 2\,\text{m}$ cooled to $T=4\,\text{K}$, and assuming a mechanical quality factor of $Q_{\rm mech} \sim 10^6$, then at a resonant frequency of $\w_{\rm mech} \sim 10^4 \,\text{Hz}$, the force PSD is $S_{F}^{\rm th} \sim 10^{-20}\, \text{N}^2/\text{Hz}$.
The above expression is approximate, as to compute the actual flux requires evaluating the average displacement due to thermal vibrations and computing the induced magnetic field using the Biot-Savart law. Instead, for computing the sensitivity, we compare displacement PSDs on resonance directly as detailed below.
For frequencies above a mechanical resonant frequency,
\be
S_{\Phi}^{\rm th. mech.} \sim 10^{-51}\,\text{Wb}^2/\text{Hz} \times \left(\frac{10^5\,\text{Hz}}{\w} \right)^4\left(\frac{B_0}{10\,\text{T}} \right)^2 \left(\frac{A_p}{\pi\times (0.4\,\text{m})^2\,} \right)^2\!,
\ee
where the values for the other parameters were provided above.
Below the mechanical resonance, $S_{\Phi}^{\rm th. mech.} \sim 10^{-47}\,\text{Wb}^2/\text{Hz}$, while on the mechanical resonance $S_{\Phi}^{\rm th. mech.} \sim 10^{-35}\,\text{Wb}^2/\text{Hz}$.
Comparing the thermal mechanical noise on resonance with the SQUID noise, we see that only for a bar mass of $M \sim 10^{11}\,\text{kg}$ (similar to an asteroid mass) would SQUID noise become the dominant noise source.

The pickup loop will also experience thermal vibrations and can therefore contribute noise.
The corresponding displacement PSD again takes the form of Eq.~\eqref{eq:mechPSDx}, albeit with modified values.
How these displacements of the loop translate to magnetic flux depends on where the loop is placed in the magnet.
For a coaxial loop near $z'=0$, the field is particularly uniform, and the variation in the magnetic field scales as $\sim B_0 (x/r_1)^2$.
Near $z'=\ell/2$, the gradient is larger and the field variations scales linearly, $\sim B_0 x/\ell$.
Let us assume the linear scaling to obtain a conservative estimate of the noise contribution using Eq.~\eqref{eq:mechPSD}.
We assume the pickup loop has a thickness $d\sim \text{mm}$, a resonant frequency $\w_p/2\pi \sim 10^6\,\text{Hz}$, a $Q_p \sim 10^3$, a weight of 1\,kg, and that it is cooled to $T = 4\,\text{K}$.
We then find that for $\w \ll \w_p$, $S_{\Phi}^{\rm th. loop} \sim 10^{-48}\,\text{Wb}^2/\text{Hz}$, for $\w \sim \w_p$, $S_{\Phi}^{\rm th. loop.} \sim 10^{-42}\,\text{Wb}^2/\text{Hz}$, and for $\w \gg \w_p$, $S_{\Phi}^{\rm th. loop.} \sim 10^{-53}\,\text{Wb}^2/\text{Hz} \times (10^8\,\text{Hz}/\omega)^4$.
For $\w \gg \w_{\rm mech}$, the thermal noise from the pickup loop can dominate that of the magnet, although it remains well below that of the SQUID.

\begin{figure}[t]
\centering
\includegraphics[width = 0.4 \textwidth]{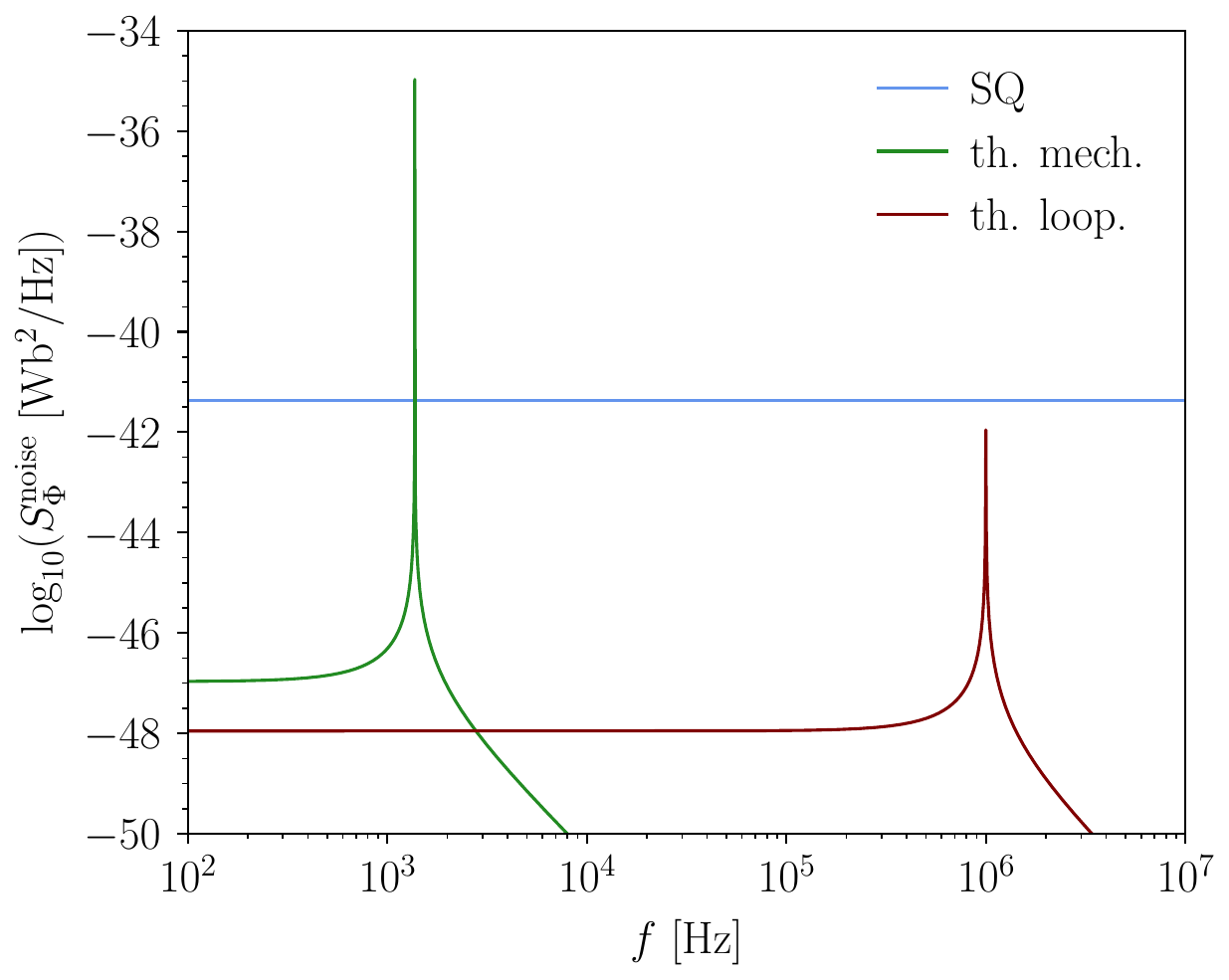}
\hspace{0.5cm}
\includegraphics[width = 0.4 \textwidth]{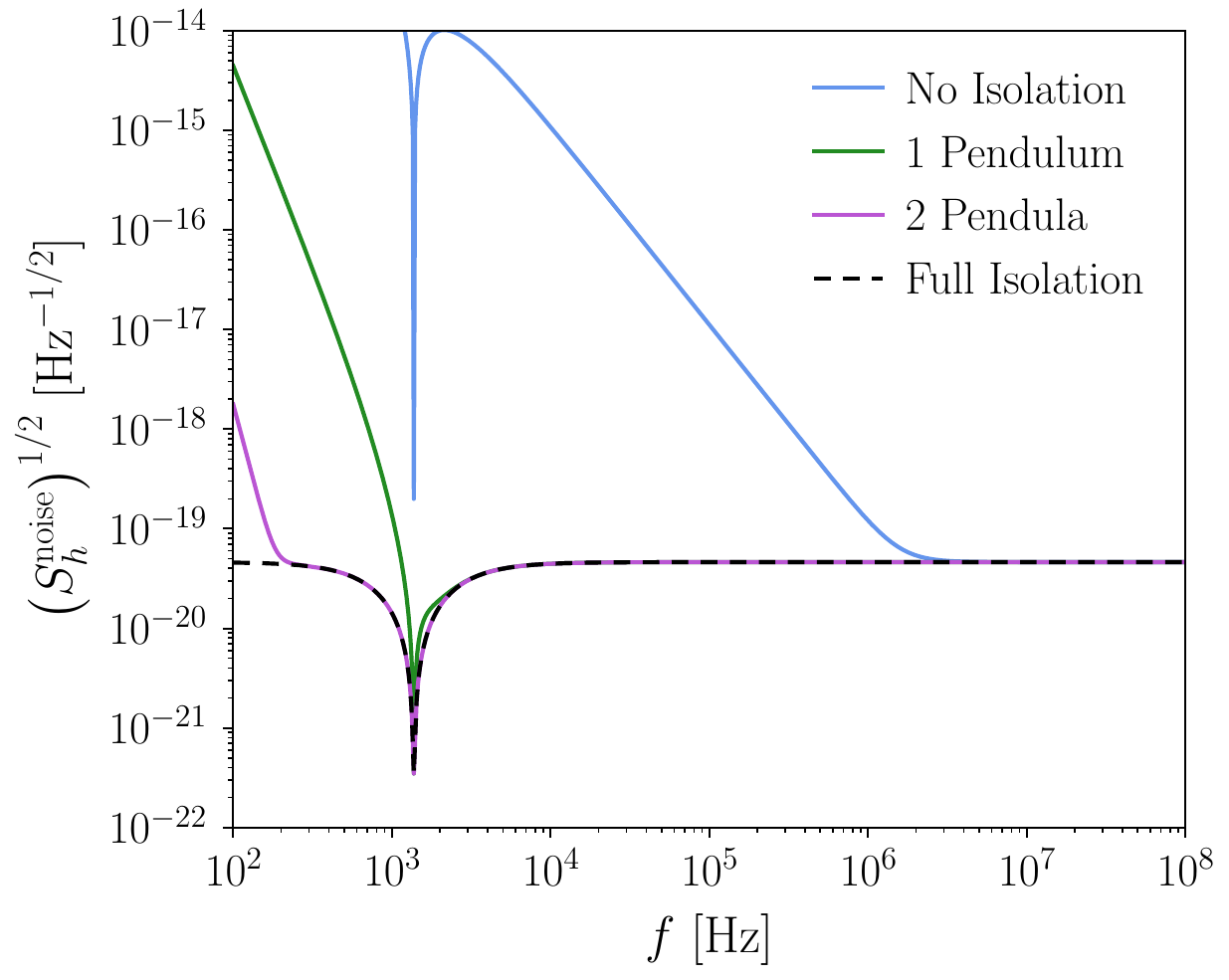}
\caption{(Left) A comparison of the PSD flux noise we expect from the SQUID (SQ blue), thermal vibrations within the magnet (th. mech. green), and thermal vibrations within the pickup loop (th. loop. red).
These are the relevant backgrounds to consider for a broadband readout, assuming the system has been seismically isolated.
Parameters adopted are discussed in the text along with further details.
(Right) An estimate of the impact of seismic noise on the ADMX-EFR sensitivity shown in Fig.~\ref{fig:ShPlotMagnet}.
There we assumed isolation with two pendula, and as can be seen the sensitivity could be degraded significantly below a MHz if no isolation at all is employed.}
\label{fig:BkgBroad}
\end{figure}

All three noise sources discussed so far are shown in Fig.~\ref{fig:BkgBroad}.
We emphasise that a more precise modelling of the magnet and loop response to thermal vibrations should be accounted for in the determination of the contribution to the noise budget in an actual experiment.

So far we have assumed that the dominant vibrational noise source is of thermal origin.
However, for a system that is not seismically well-isolated, environmental seismic noise can dominate.
A typical displacement PSD at a relatively quiet site yields~\cite{Saulson:2017jlf}
\be
S^{\rm seismic}_x(\w) \simeq 10^{-14} \text{cm}^2/\text{Hz} \times \text{min}\left[1,\, \left(\frac{ 10\,\text{Hz}}{\omega/2\pi}\right)^4\right]\!.
\ee
Comparing this with the thermal mechanical displacement PSD, we see that on the mechanical resonant frequency, $S^{\rm seismic}_x \simeq 60 S_x^{\rm th.}$.
This highlights the need for at least some form of seismic isolation.
An example of a passive isolation system would be to suspend both the magnet and the pickup loop by a cable whose natural frequency is $f_{\rm pend.} \sim 2\,\text{Hz}$, which would suppress the seismic displacement PSD by a factor $(2\,\text{Hz}/f)^4$, making seismic noise significantly sub-dominant to thermal vibration noise near the mechanical resonances.
In Fig.~\ref{fig:BkgBroad} we show the impact on our projected ADMX-EFR sensitivity of having no seismic isolation at all as compared to having one or two pendula and total isolation.
For Fig.~\ref{fig:ShPlotMagnet} in the main text, we assumed isolation with two pendula.
Even with a suspension system, at sufficiently low frequencies below the mechanical resonance gravity gradient noise must be accounted for~\cite{Saulson:1984yg,Hughes:1998pe}.

The noise-equivalent strain PSD plotted in Fig.~\ref{fig:ShPlotMagnet} in the main text is obtained from Eqs.~\eqref{eq:sigPSD},~\eqref{eq:mechPSD} and~\eqref{eq:squidPSD} as
\be
S_h^{\rm noise}(\w) = \frac{S_{\Phi}^{\rm th. mech.}(\w) + S_{\Phi}^{\SQ}(\w)}{S_{\rm sig}(\w) /S_h(\w)}.
\ee
When $\w\sim \w_{\rm mech}$, this leads to mechanical thermal noise dominating. 
In order to compute the noise-equivalent strain PSD on the mechanical resonance, we do not make use of the approximate expression given in Eq.~\eqref{eq:mechPSD}.
Instead, we make use of the fact that the thermal force PSD leads to the displacement PSD given in Eq.~\eqref{eq:mechPSDx}, which can be directly compared with the expected displacement PSD due to the GW force, $S_x^{h}(\w_{\rm mech}) \simeq \left(Q_{\rm mech} \eta_{210} \ell_\eta \bu_{210} \right)^2 S_h(\w_{\rm mech})$.
Since both the thermal force and GW force couple dominantly to the 210 mode in the vicinity of that mode's resonance, the direct comparison of the displacement PSDs allows for a computation of $(S_h^{\rm noise})_{\rm mech}$ without the need to explicitly compute the flux (and therefore gain) due to the thermally-induced displacements.
Performing this comparison of the displacement PSDs, we obtain on resonance
\be
\left(S_h^{\rm noise}\right)_{\rm mech} \simeq \frac{2 T}{M (\eta_{210}\ell_\eta \bu_{210})^2 \w_{\rm mech}^3 Q_{\rm mech}}.
\label{eq:Sh_mech_res}
\ee
In this limit there is no dependence on EM properties of the magnet and the scaling is identical to that of a conventional Weber bar.
Our sensitivity compared to AURIGA in Fig.~\ref{fig:ShPlotMagnet} can be entirely attributed to factors in Eq.~\eqref{eq:Sh_mech_res}, most notably the larger mass of the magnets and for DMRadio the reduced temperature.
Away from a mechanical resonance, SQUID noise dominates, such that we obtain
\be
\left(S_h^{\rm noise}\right)_{\SQ} \simeq \frac{4 L_p}{\alpha^2 L\,B_0^2\,A_{p}^2}\frac{10^{-12}\Phi_0^2}{\text{Hz}},
\label{eq:NES-SQ}
\ee
where we have assumed $|\mathcal{G}|^2 = 1$.

\subsection{Resonant LC circuit noise}

The above noise sources are relevant to a broadband setup. It is also possible to couple the pickup loop to a resonant EM detector, such as an LC circuit as planned by the DMRadio collaboration.
In this scenario, the sensitivity could benefit from the EM quality factor when operating on the EM resonance.
However, some noise sources are also filtered by the resonant response, so we must rescale our noise PSDs to account for this coupling.
All fluxes seen by the SQUID are rescaled with respect to the flux seen by the pickup loop by a factor~\cite{Foster:2017hbq}
\be
\Phi_{\rm res}^{\SQ} = \alpha Q_{\EM} \sqrt{\mathcal{T}(\w)} \frac{\sqrt{L L_i}}{L_p+L_i} \Phi^{\rm loop},
\ee
where $L_i$ is the inductance of the resonant circuit, and the transfer function is
\be
\mathcal{T}(\w) \equiv \frac{\w^4/Q_{\EM}^2}{(\w_{\EM}^2 - \w^2)^2 + (\w_{\EM} \w/Q_{\EM})^2},
\ee
for an EM resonant frequency $\w_{\EM}$ with quality factor $Q_{\EM}$.
This rescaling affects the SQUID and thermal mechanical noise sources previously considered.

Operating in resonant mode adds thermal noise from dissipation in the LC circuit.
The resultant noise PSD is~\cite{Foster:2017hbq}
\be
S_{\Phi}^{\rm th.\,LC} = 2\alpha^2 Q_{\EM} \mathcal{T}(\w) \frac{\w_{\EM}}{\w^2} \frac{L L_i}{L_p+L_i} \, T_{\LC},
\ee
where $T_{\LC}$ is the system temperature.
On the EM resonance and with $L_i = L_p$, we can evaluate this as
\be
S_{\Phi}^{\rm th.\,LC} \sim 10^{-32}\,\text{Wb}^2/\text{Hz} \times \left(\frac{10^5\,\text{Hz}}{\w_{\EM}} \right) \left( \frac{Q_{\EM}}{2 \times 10^7} \right) \left(\frac{T_{\LC}}{10\,\text{mK}} \right)\!.
\ee
The system temperature and $Q_{\EM}$ are taken to match the DMRadio-GUT projection~\cite{DMRadio:2022jfv}.
However, on resonance the signal PSD is enhanced by a factor of $Q_{\EM}^2 = 4 \times 10^{14}$, making the benefit of going to a resonant detector clear.

The thermal noise in the LC circuit clearly swamps the SQUID noise, as expected.
On the mechanical resonance, however, the magnet vibrational noise is resonantly enhanced both mechanically and electromechanically, such that it remains the dominant noise source and there is no advantage to using an LC resonator near $\w \sim \w_{\rm mech}$.
At higher frequencies, however, there is an advantage, and the noise-equivalent strain PSD scales as
\be
\left(S_h^{\rm noise}\right)_{\LC}^{\w \gg \w_{\rm mech}} \simeq \frac{4 L_p T_{\LC}}{Q_{\EM} \w_{\EM} B_0^2 A_p^2}.
\ee
In this limit, the improvement over the broadband reach in Eq.~\eqref{eq:NES-SQ} is
\be
\frac{Q_{\EM} \omega S_{\Phi}^{\SQ}}{\alpha^2 T_{\LC} L} \simeq 10^5 \left( \frac{\omega/2\pi}{10^4\,\text{Hz}} \right)\!,
\ee
as claimed in the main text.

\section{Comparison with existing approaches}

Finally, we return to consider in more detail the question of why our proposal of using a magnet as a Weber Bar could achieved enhanced GW sensitivity beyond a conventional Weber Bar (also referred to as a resonant mass detector) or approaches where the GW couples directly to the magnetic field, as can be inferred from Fig.~\ref{fig:ShPlotMagnet}.
We discuss each in turn below.

\subsection{Canonical Weber Bars}

In a typical Weber bar experiment, the GW modulates the stored elastic energy in the resonator, given by
\be
U_{\rm elastic} = \frac{1}{2} M \w^2 x^2,
\ee
where $x$ is the displacement that will be linear in the GW strain.
Assuming a flexible detector such that $\w \gg \w_{\rm mech}$, with $\w_{\rm mech}$ the resonant frequency of the bar, the response function of $x$ to a GW is such that the stored elastic energy modulated by the GW is $U_{\rm elastic}^h \sim h^2 U_{\rm elastic}$.
Typical bars often weigh 1000s of kgs, corresponding to a total accessible energy that can be enormous,
\be
U_{\rm elastic}^h \sim 10^{12}\,\text{J} \times h^2 \,\left( \frac{\w}{2\pi\times10\,\text{kHz}}\right)^2\left(\frac{M}{1000\,\text{kg}} \right) \left(\frac{L}{1\,\text{m}} \right)^2\!,
\ee
which is far larger than the electromagnetic energy stored in e.g. the ADMX-EFR magnet, which is of the order
\be
U_{\rm magnet}^h \sim 10^8 \,\text{J} \times h^2 \left(\frac{L}{1\,\text{m}} \right)^3 \left(\frac{B_0}{10\,\text{T}} \right)^2\!.
\ee
Not only is the EM energy smaller, but it is independent of frequency, while the elastic energy is growing at larger frequencies.
This would suggest, then, that a traditional Weber bar could be better than our not-so-static magnet.

However, two important points affect the comparison.
The first is the way in which a Weber bar's length modulation is measured by converting elastic energy into EM energy in a transducer (often resonant).
If we were to take a simple setup of a metallic bar, one end of which acts as one of the two plates in a parallel-plate capacitor, we have a set of two coupled equations for the system~\cite{Vinante:2006uk}.
In the limit of no EM backaction on the bar, we effectively have the following energy flow: GW energy is transferred into the elastic energy of the bar, which then gets converted into EM energy in the capacitor. 
Usually, a DC electric field of magnitude $E_C \sim 10\,\text{MV/m}$ is applied across the plates, and the voltage therefore scales as $V_C \sim E_c (h L)$. As a result, the EM energy transferred to the capacitor is
\be
U_{\rm EM,\,C} = \frac{1}{2}C\, V_C^2 \sim 5\times 10^5\,\text{J} \times h^2 \left(\frac{L}{1\,\text{m}} \right)^2 \left(\frac{E_C}{10\,\text{MV/m}} \right)^2 \left(\frac{C}{10^{-8}\,\text{F}}\right)\!,
\ee
where we have taken typical values for e.g. the AURIGA resonant bar~\cite{Vinante:2006uk}.
We immediately see that the total EM energy in the capacitor is actually much smaller than that of the magnet, primarily due to the fact that it is much easier to generate large magnetic fields than electric fields in the laboratory.
The smallness of this capacitor energy with the elastic energy above tells us that our no backaction limit is a reasonable assumption.

The second important point to consider when comparing traditional Weber bars with our proposed magnet experiment is that both are actually subject to the same limiting behaviour when the GW frequency matches a mechanical resonant frequency.
The reason for this is that on a resonance, mechanical thermal noise typically dominates over EM noise (SQUID or LC thermal), such that the EM stored energy factors out in the SNR (see Eq.~\eqref{eq:SNR_mech_res} in the main text).
The result is that in this regime the SNR depends on the mechanical stored energy, making different Weber bars of similar mass and dimension comparable.
However, if the mass of the bar is increased to suppress thermal mechanical noise below EM noise, it is once again the EM stored energy which dictates the sensitivity of the apparatus.

The conclusion we reach here is that the ceiling for elastic GW detectors is not yet attained, since there is far more elastic energy being stored than there is EM energy, both in a typical Weber bar experiment and in our magnet proposal.
When searching for a signal degenerate with a mechanical resonance, if the elastic energy has not been increased to the point where EM noise dominates over thermal vibration noise, then the ultimate Weber bar sensitivity has not been reached.
However, away from the mechanical resonance, the much larger EM field in our magnet proposal makes ours a more efficient Weber bar than previous Weber bar experiments.

\subsection{Electromagnetic GW Coupling}

Consider next a direct comparison with strategies that exploit the coupling of a GW to an electromagnetic field directly.
That approach involves searching for the effect of a coupling of the GW to $B_0$ through an interaction term $\sim h F^2$~\cite{Ejlli:2019bqj,Berlin:2021txa,Domcke:2022rgu}, which as has been studied with analogy to the search for axion dark matter.
For the frequency range we consider, the most appropriate comparison would be to the resonant LC strategy discussed in Ref.~\cite{Domcke:2022rgu}, which envisioned using the DMRadio detector as a GW telescope.
As shown in Fig.~\ref{fig:ShPlotMagnet}, that strategy can have a comparable noise-equivalent strain sensitivity at the highest frequencies, however, it degrades rapidly at lower frequencies.
(We emphasise again that our sensitivity is for a broadband readout, whereas the resonant LC reach is for a resonant sensitivity.)
Our main focus here is to review why a different scaling with frequency occurs for a mechanical coupling, although we refer to Ref.~\cite{Berlin:2023grv} for a more detailed discussion of the comparative frequency scaling of various proposals.

We can understand the key difference in behaviours as follows.
Working in the proper detector frame of the magnet, the GW induces a spacetime curvature, and therefore an effect second order in frequency, $\sim \w^2 h$ (see e.g. Ref.~\cite{Domcke:2023bat}).
The GW acts to excite the magnet mechanically and subsequently the mechanical oscillations excite an electromagnetic mode of the system.
The resulting magnetic field that is sourced must depend on two transfer functions,
\be
B_h \sim  h B_0\frac{\w_{\EM}^2 }{(\w^2-\w_{\EM}^2)}\frac{\w^2}{(\w^2-\w_{\rm mech}^2)},
\label{eq:Bh-omegascaling}
\ee
where we neglect the finite width near the resonances.
Parametrically, the electromagnetic resonant frequency is $\w_\EM \sim 1/\ell$ and therefore ${\cal O}(100\,{\rm MHz})$ for a $1\,{\rm m}$ device.
The mechanical resonance will be suppressed by the sound speed $c_s \sim 10^{-5}$ (an appropriate value for most materials), so that $\w_{\rm mech} \sim c_s \w_{\EM} \ll \w_{\EM}$.
For $\w_{\rm mech} \ll \w \ll \w_{\EM}$, $B_h$ is independent of frequency, whereas when $\w \ll \w_{\rm mech}$, $B_h \propto \w^2$ and the sensitivity degrades quadratically at low frequencies.
If we couple the GW to the magnetic field directly, then the analog of Eq.~\eqref{eq:Bh-omegascaling} would be
$B_h \sim  h B_0\w^2/(\w^2-\w_{\EM}^2)$, so that for $\w \ll \w_{\EM}$ the sensitivity degrades as $B_h \propto \w^2$.
(These estimates cannot be extended above $\w_{\EM}$ as the signal the GW sources will no longer be coherent over the detector volume, which generically necessitates a different detector configuration~\cite{Benabou:2022qpv}.)

Lastly, note that extrapolated to the EM resonance of the detector at ${\cal O}(100\,\text{MHz})$ there is no advantage to our proposal over measuring the EM coupling through a resonant LC circuit.
The reason the appropriate curves in Fig.~\ref{fig:ShPlotMagnet} would not cross if extrapolated to that scale is due to the enhanced sensitivity of DMRadio-GUT, compared to what is assumed for the resonant LC curve.

\end{document}